%% file: ms.tex
\documentclass[revtex4]{emulateapj}

\usepackage{graphicx}
\usepackage{epstopdf}
\usepackage{color}
\usepackage{amsmath}
\usepackage{xspace}
\allowdisplaybreaks[1]
\newcommand{\mbh}{\ensuremath{M_{\mathrm{BH}}}\xspace}
\newcommand{\msigma}{\ensuremath{M_{\mathrm{BH}}-\sigma}\xspace}
\newcommand{\mlum}{\ensuremath{M_{\mathrm{BH}}-L}\xspace}
\newcommand{\msun}{\ensuremath{M_\odot}\xspace}

\newcommand{\teff}{\ensuremath{T_{\mathrm{eff}}}\xspace}
\newcommand{\kband}{\ensuremath{K} band\xspace}
\newcommand{\kbandadj}{\ensuremath{K}-band\xspace}
\newcommand{\zband}{\ensuremath{Z} band\xspace}
\newcommand{\zbandadj}{\ensuremath{Z}-band\xspace}
\newcommand{\kms}{\ensuremath{\mathrm{km}~\mathrm{s}^{-1}\xspace}}

\begin{document}
\title{Prospects for measuring supermassive black hole masses with future extremely large telescopes}
\author{Tuan Do\altaffilmark{1,8}, Shelley A. Wright\altaffilmark{1,2},  Aaron J. Barth\altaffilmark{3}, Elizabeth J. Barton\altaffilmark{3}, Luc Simard\altaffilmark{4},  James E. Larkin\altaffilmark{5}, Anna M. Moore\altaffilmark{6}, Lianqi Wang\altaffilmark{7}, and Brent Ellerbroek\altaffilmark{7}} 
\altaffiltext{1}{Dunlap Institute for Astronomy and Astrophysics,
University of Toronto, 50 St. George Street, Toronto M5S 3H4, ON, Canada}
\altaffiltext{2}{Department of Astronomy and Astrophysics,
University of Toronto, 50 St. George Street, Toronto M5S 3H4, ON, Canada}
\altaffiltext{3}{Department of Physics and Astronomy, 4129 Frederick Reines Hall, University of California, Irvine, CA 92697-4575, USA}
\altaffiltext{4}{Herzberg Institute of Astrophysics, National Research Council of Canada, Victoria, BC, V9E 2E7, Canada}
\altaffiltext{5}{Physics and Astronomy Department, University of California,
    Los Angeles, CA 90095-1547}
\altaffiltext{6}{Caltech Optical Observatories, California Institute of Technology, Pasadena, CA}
\altaffiltext{7}{TMT Observatory Corporation Instrumentation Department, Pasadena, CA}
\altaffiltext{8}{Dunlap Fellow}
\begin{abstract}
The next generation of giant-segmented mirror telescopes ($>$ 20 m) will enable us to observe galactic nuclei at much higher angular resolution and sensitivity than ever before. These capabilities will introduce a revolutionary shift in our understanding of the origin and evolution of supermassive black holes by enabling more precise black hole mass measurements in a mass range that is unreachable today. We present simulations and predictions of the observations of nuclei that will be made with the Thirty Meter Telescope (TMT) and the adaptive optics assisted integral-field spectrograph IRIS, which is capable of diffraction-limited spectroscopy from $Z$ band (0.9 $\micron$) to $K$ band (2.2 $\micron$). These simulations, for the first time, use realistic values for the sky, telescope, adaptive optics system, and instrument, to determine the expected signal-to-noise ratio of a range of possible targets spanning intermediate mass black holes of $\sim10^4$ \msun to the most massive black holes known today of $>10^{10}$ $M_\odot$. We find that IRIS will be able to observe Milky Way-mass black holes out the distance of the Virgo cluster, and will allow us to observe many more brightest cluster galaxies where the most massive black holes are thought to reside. We also evaluate how well the kinematic moments of the velocity distributions can be constrained at the different spectral resolutions and plate scales designed for IRIS. We find that a spectral resolution of $\sim8000$ will be necessary to measure the masses of intermediate mass black holes. By simulating the observations of galaxies found in SDSS DR7, we find that over $10^5$ massive black holes will be observable at distances between $0.005 < z < 0.18$ with the estimated sensitivity and angular resolution provided by access to $Z$-band (0.9 $\micron$) spectroscopy from IRIS and the TMT adaptive optics system. These observations will provide the most accurate dynamical measurements of black hole masses to enable the study of the demography of massive black holes, address the origin of the $M_{\mathrm{BH}}-\sigma$ and $M_{\mathrm{BH}}-L$ relationships, and evolution of black holes through cosmic time. 
\end{abstract}


\section{Introduction}

Mass measurements of black holes in galactic nuclei over the past decade have shown they are a fundamental byproduct of galaxy evolution. The observed relationships between black hole mass and galaxy properties (e.g. the velocity dispersion, luminosity of the bulge/spheroid, and mass of the bulge/spheroid) suggest that galaxies and their central black holes co-evolve and galaxy growth may be regulated by feedback from their central black holes \citep[e.g.][and references therein]{1995ARA&A..33..581K,1998AJ....115.2285M,2000ApJ...539L...9F,2000ApJ...539L..13G,2012MNRAS.419.2497B}. Understanding the nature of these scaling relationships relies observationally on obtaining accurate black hole mass measurements, as well as building up a larger statistical sample to understand potential deviations from differing galaxy samples, environment, and evolution. Among the most robust methods of black hole mass measurements involve using dynamical tracers such as the orbits of water masers or through stellar dynamics. These dynamical measurements are further important to calibrate other methods of measuring black hole masses, such as reverberation mapping and translating the measurements of AGN line widths to black hole masses in the distant universe \citep[e.g.][]{2006ApJ...646..754D,2007ApJ...670..105O}. In recent years, stellar dynamical measurements within the black hole's sphere of influence have led to accurate black hole masses for $\sim70$ sources, largely in the range of $10^7 - 10^9$ \msun \citep[e.g.][]{2009ApJ...698..198G,2010MNRAS.401.1770V,2011ApJ...729...21S,2013ApJ...764..184M} with only a few of the most massive black holes at $10^{10}$ \msun \citep[e.g.][]{2011Natur.480..215M,2012ApJ...756..179M}. Many of black hole mass measurements are from the STIS instrument on the \textit{Hubble Space Telescope} (\textit{HST}), which provides both the necessary spatial and spectral resolution to resolve the sphere of influence of black holes in this range of masses. With the advent of integral field spectrographs (IFS) behind adaptive optics (AO), it is now possible to obtain stellar dynamical measurements of black hole masses from the ground on 8-10 m telescopes \citep[e.g.][]{2009MNRAS.399.1839K,2011ApJ...729..119G,2011MNRAS.410.1223R,2011ApJ...728..100M,2012ApJ...753...79W}. The IFS measurements have the unique advantage of being able to measure both the spatial and spectral dimensions simultaneously. This allows for an efficient way of sampling spatially dependent stellar kinematics for the dynamical models that are used to determine black hole masses. Yet, even with AO and large telescopes, the range of measurable black hole masses is limited by the angular resolution and sensitivity of telescopes today. 

While substantial progress in observational measurements of black hole masses has been made, there are a number of critical questions that are not possible to investigate with current capabilities. These difficulties stem from our lack of knowledge about the high mass ($>10^9$ \msun) and low mass ($< 10^7$ \msun) ends of the scaling relationships. For example, at the high mass end of the \msigma and \mlum relationships, the predictions for black hole mass from these two relationships begin to diverge significantly; at $M_{BH} \approx 10^9$ \msun, the \mlum relationship predicts a larger black hole mass than the \msigma relationship as the velocity dispersion of these massive galaxies stop increasing with galaxy luminosity \citep[e.g.][]{2007ApJ...670..249L,2011Natur.480..215M}. These discrepancies have important implications for the space density of massive black holes in the universe. If the \mlum relationship is more fundamental than the \msigma, it would predict over an order of magnitude more $> 10^9$ \msun black holes \citep{2007ApJ...662..808L}. The space density of very massive black holes is not well constrained because of the difficulty detecting many of these black holes given the distance to these massive galaxies and the sensitivity and resolution of current telescopes. 

Much about the scaling relationships between central black hole mass and galaxy properties are even unknown at the well-sampled $10^{7} \msun < \mbh < 10^9$ \msun range. For instance, investigating various galaxy samples yield differing scaling relationships between late and early type galaxies \citep{2013ApJ...764..184M}; this is further complicated with observational biases from distinct black hole mass measurement methods (e.g., stellar, gas, or masers). Indeed, the observations of black holes with masses less than $< 10^6$ \msun, or intermediate mass black holes (IMBHs), are difficult at best and have been subject to conflicting interpretations. For example, reports of a central black hole with M$_{BH} \sim 4\times10^4$ \msun in the Omega Centauri globular cluster using radial velocity measurements \citep{2008ApJ...676.1008N} have been disputed by subsequent observations of proper motions with \textit{HST} \citep{2010ApJ...710.1032A}. The measurement a black hole with M$_{BH} \sim2\times10^4$ \msun in the G1 globular cluster in M31 \citep{2002ApJ...578L..41G}  may also be consistent with a kinematic model of the cluster without a black hole \citep{2003ApJ...589L..25B}. These measurements are disputed because they are made at angular resolutions comparable to the scale of the black hole's gravitational sphere of influence, which results in measurements that are strongly influenced by the physical properties of the stellar light distribution that contributes to the kinematic measurement. A steep mass density profile of the stars for example, can cause an apparent increase in the velocity dispersion at the center that, if not properly accounted for, will bias the black hole mass measurement. The precision in the moments of the line of sight velocity distribution (LOSVD) is also crucial for obtaining accurate black hole masses. Because of the low masses of IMBHs and their hosts, the velocity dispersions at the radius of influence (within which the gravitational influence of the black hole dominates) are expected to be only a few tens of km s$^{-1}$. In order to obtain a precision of $\approx 1\times10^4$ \msun, the uncertainties in velocity dispersion measurement must be on the order of 1-3 km s$^{-1}$. A spectral resolution of $R > 5000$ is necessary to reach such precision in the velocity dispersion and to reliably measure the higher velocity moments. Increasing our sensitivity to  black hole mass measurements is crucial to determine the minimum mass of central black holes. These limits would help differentiate between the two main theories for the origin of supermassive black holes: (1) massive black holes form as the byproduct of the collapse of numerous massive stars, or (2) the black holes start as the collapse of a massive gas cloud directly into a central black hole. The first scenario would predict a large number of intermediate mass black holes today, while the second would predict a dearth of these lower mass black holes today \citep[e.g.][]{2010A&ARv..18..279V}.

\begin{figure}[th!]
\centering
\includegraphics[angle=90,width=3.4in]{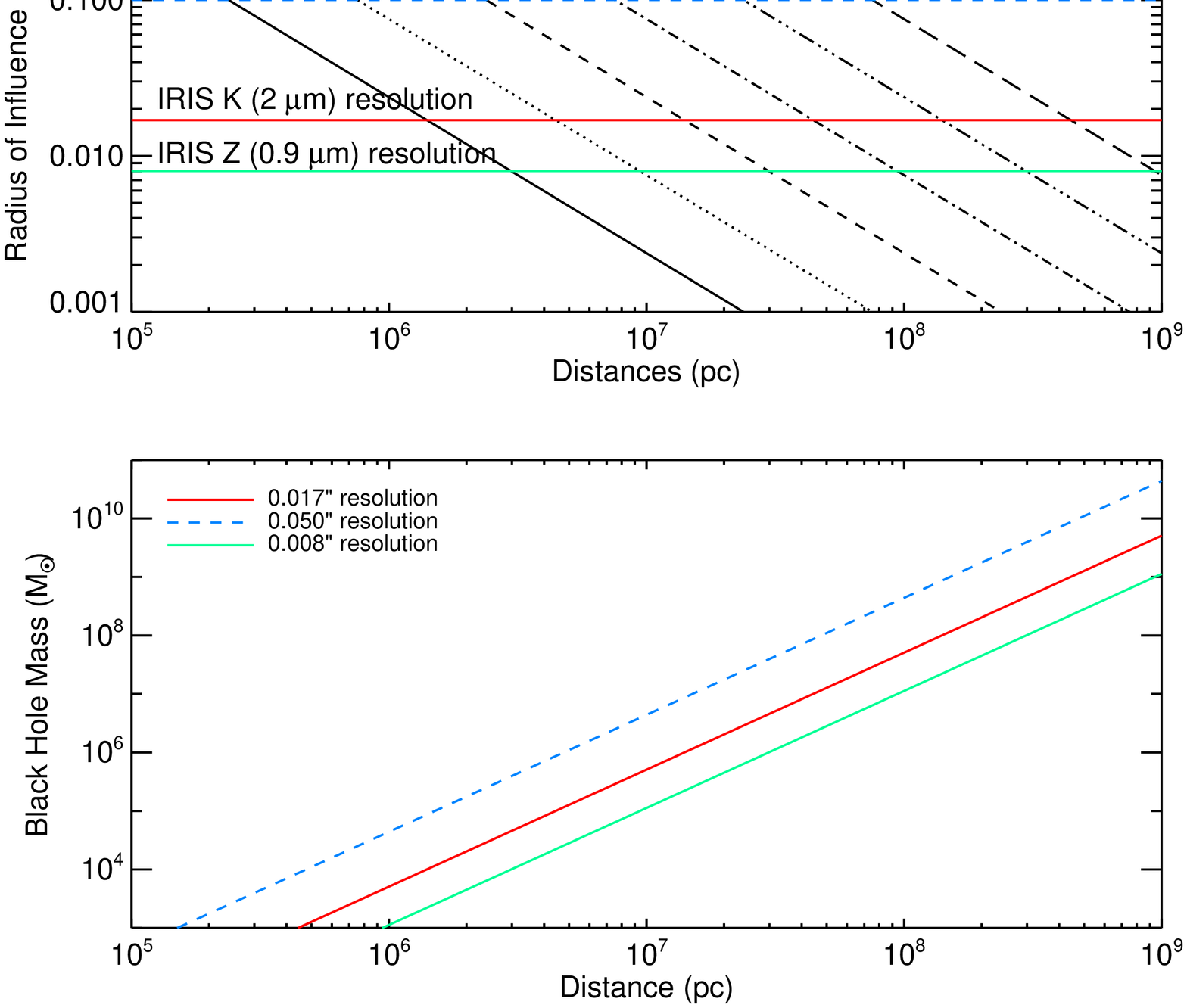}
\caption{\textbf{Top:} the projected radius of influence in the plane of the sky for various black hole masses as a function of the angular size distance using the observed \msigma relationship \citep{2009ApJ...698..198G}. The lines for black hole masses, increasing from left to right, from $10^4$ $M_\odot$ (solid) to $10^9$ $M_\odot$ (dashed). The increase in angular resolution will allow observations of black holes much further than possible today. \textbf{Bottom:} the lower limits on black hole mass measurements as function of distance. Most IFS BH mass measurements today are made at a pixel scale of 50 mas (100 mas resolution; blue, dashed), compared to future GSMTs, which will have angular resolution of 18 mas at \kband (red, solid) and 8 mas at \zband (green, solid).}
\label{fig:bh_influence}
\end{figure}

One of the goals of the next generation of giant segmented mirror telescopes (GSMT) is to provide the angular resolution and sensitivity to answer these questions about the origin and evolution of supermassive black holes. A simple way to compare the capabilities of current 8-10 m telescopes for measuring black hole masses to that of GSMTs is through how well the gravitational sphere of influence of the black holes can be spatially resolved. The radius of influence is defined as the location from a black hole within which the potential from the black hole dominates \citep{2008gady.book.....B}:
\begin{equation}
r_{\textrm{\footnotesize infl}} = \frac{G \mbh}{\sigma},
\end{equation}
where $\mbh$ is the black hole mass and $\sigma$ is the velocity dispersion at the effective radius, as measured in the $\msigma$ relationship. To remove the dependence on $\sigma$, we use the \msigma relationship $\mbh \propto \sigma^4$ \citep{2000ApJ...539L...9F,2000ApJ...539L..13G,2009ApJ...698..198G}.
Figure \ref{fig:bh_influence} shows the angular radius of influence of various mass black holes as a function of physical distance; the increase in angular resolution translates to a larger range of black hole masses that are accessible at greater distances. While it is clear that future GSMTs with diameters of about 30 m equipped with AO systems will provide a revolutionary increase in angular resolution, little work have been done to quantitatively predict the quality and and limitations of the data that will be used to measure black hole masses. In particular, we need to assess the \textit{observability} of the nuclei of galaxies of interest using our current knowledge of their light profiles as well as modeling the noise sources involved such as the sky background, telescope, and instrument. 

As a concrete example of the power of future instruments on GSMTs, we develop a data simulator for the first-light AO fed integral-field spectrograph, the InfraRed Imaging Spectrograph (IRIS), for the Thirty Meter Telescope (TMT). In particular the simulations are used to assess the signal to noise ratio (SNR) attainable by the IRIS spectrograph using realistic values for the expected instrument and telescope performance for a set of representative objects in which a black hole mass measurement would be of scientific interest. We present details about the instrument in Section \ref{sec:instrument} and the data simulator in Section \ref{sec:simulation}. In Section \ref{sec:gauss_hermite}, we also investigate the relationship between SNR and the precision of kinematic moment measurements that will ultimately be used in dynamical mass modeling. In Section \ref{sec:results}, we present the results of simulations of both massive early type galaxies such as those from \citet{2007ApJ...664..226L}  as well as lower mass systems such as the Milky Way-like black holes and the G1 globular cluster to assess the whole range of black hole masses accessible in the era of 30 m class telescopes. We estimate, based on bulged/disk decompositions of galaxies from SDSS DR7, the number of galaxies that will be observable with IRIS and TMT in Section \ref{sec:sdss}. The simulated capabilities of IRIS/TMT are compared to the current state of IFS measurement of black hole masses in Section \ref{sec:current}. In Section \ref{sec:conclusion}, we present our conclusions.

\section{Description of the IRIS instrument}
\label{sec:instrument}
The InfraRed Imaging Spectrograph \citep[IRIS,][]{2010SPIE.7735E..76L} is designed to be one of the first light instruments for the Thirty Meter Telescope (TMT), with both imaging and integral-field spectroscopy capabilities. It will be fed by an AO system, the Narrow-Field Infrared Adaptive Optics System \citep[NFIRAOS,][]{2010SPIE.7736E...9H}, to provide diffraction limited resolution. IRIS will be a workhorse instrument, spanning science cases from solar system objects to high-z galaxies, but here we will summarize its capabilities that are relevant to dynamical measurement of black hole masses. IRIS has both an image slicer and a lenslet spectrograph to provide four spatial pixel (spaxel\footnote{Following common usage in IFS observations, we will use the term spaxel to mean a spatial pixel in the IFS data cubes and spectral channel to mean a pixel in the wavelength dimension.}) scales - 4 mas, 9 mas, 25 mas, and 50 mas - with wavelength coverage from Z ($\lambda_\mathrm{cen} = 0.93$ $\micron$) to \kband ($\lambda_\mathrm{cen} = 2.18$ $\micron$). The lenslet mode covers the 4 and 9 mas plate scales while the image slicer uses the 25 or 50 mas scale. The field of view depends on the spatial pixel scale as well as the number of spectral elements sampled. The image slicer has $90\times45$ spaxels while the lenslet spectrograph has a $16\times128$ spaxel mode for broad-band wavelength coverage (20\% bandpass) or $112\times 128$ spaxels for more narrowband (5\% bandpass); the field of view ranges from $4.\arcsec 4 \times 2.\arcsec 25$ for the 50 mas plate scale with the 2000 spectral channels to $0.\arcsec 064\times0.\arcsec 51$ for the smallest spatial pixel scale (4 mas) with 4096 spectral channels. The most relevant mode for dynamical black hole mass measurement will likely be the 5\% bandpass filter with either the 4 or 9 mas spatial pixel scale, which will provide a field of view of $0.\arcsec 45 \times 0.\arcsec 51$ and $1.\arcsec 01 \times 1.\arcsec 15$, respectively (see Table \ref{tab:fov}). The complement of broad-band filters as well as total instrument and telescope throughput in each band-pass is given in Table \ref{tab:filters}. These modes will provide spectral resolution of 4000, however a spectral resolution of 8000 will also will be possible for 2.5\% bandpass filters, which will be useful for measuring nuclei with low velocity dispersion ($< 30$ km s$^{-1}$). For galaxies with low surface brightness and black holes with large gravitational radius of influence in the plane of the sky, the 50 mas plate scale utilizing the image slicer will provide high signal-to-noise (SNR) measurements per spaxel. 

\begin{deluxetable*}{ccccc}
\tablecolumns{5}
\tablecaption{Field of view and configurations for sample IRIS observing modes}
\tablewidth{0pc}
\tabletypesize{\scriptsize}  
\tablehead{\colhead{Mode} & \colhead{Plate Scale} & \colhead{Field of View} & \colhead{Spectral Resolution} & \colhead{Bandpass} \\
           \colhead{} & \colhead{(mas)} & \colhead{(\arcsec)} & \colhead{}  & \colhead{}}
\startdata
Direct Imager & 4 & $ 16.4\times16.4$ & \nodata \\
Image Slicer ($90\times45$ spaxels) & 50 & $4.4\times2.25$ & 4000, 8000 & 20\%, 10\% \\
 & 25 & $2.2\times1.125$ & 4000, 8000 & 20\%, 10\% \\
Lenslet Spec. ($112\times128$ spaxels) & 9 & $1.01\times1.15$ & 4000 & 5\% \\
             & 4 & $0.45 \times 0.51$ & 4000 & 5\% \\
Lenslet Spec. ($16\times128$ spaxels) & 9 & $0.144\times1.15$ & 8000 & 20\% \\
             & 4 & $0.064 \times 0.51$ & 8000 & 20\% \\
\enddata
\label{tab:fov}
\end{deluxetable*}

\section{Simulation Setup}
\label{sec:simulation}
Most of the simulations of the expected SNR of IRIS observations are performed in the \kband, which is currently used for most dynamical black hole mass determinations from the ground. The \kband contains very strong CO absorption lines at $\sim 2.29-2.39$ $\micron$, which can be effectively used to determine the integrated stellar line-of-sight velocity distribution (often parameterize by the Gauss-Hermite polynomials). At these longer wavelengths, observations will also have higher Strehls from better AO correction. On current 8-10 m telescopes, the typical Strehl ratios obtained are 0.25 at $H$ band and 0.35 at \kband \citep{2006PASP..118..310V}. In comparison, the adaptive optics system for TMT will be able to achieve on-axis Strehl ratios of about 0.62 at $H$ band and 0.72 at \kband at zenith.  IRIS will be able to obtain AO-corrected observations for wavelengths as short as \zband ($\lambda = 0.840-1.026$ $\micron$) thereby covering the Ca II triplet (at $0.8498$ $\micron$, 0.8542 $\micron$, and $0.8662$ \micron), which has also been used extensively to obtain dynamical measurements of black hole masses with STIS on \textit{HST}. While the predicted Strehl ratio of 0.19 is lower than that of \kband, there is lower sky background and higher spatial resolution possible at \zband. We discuss the trade off between using the \kband, $H$ band, and \zband briefly in Appendix \ref{append:zbb}. For the bulk of the paper, we will discuss observations at \kband to best compare with current IFS observations. Filter information and sky backgrounds in each filter are giving in Table \ref{tab:filters}. \citet{2010SPIE.7735E.250W} also contains details of the IRIS simulations. 

\begin{figure*}[ht]
\centering
\includegraphics[width=2.5in,angle=90]{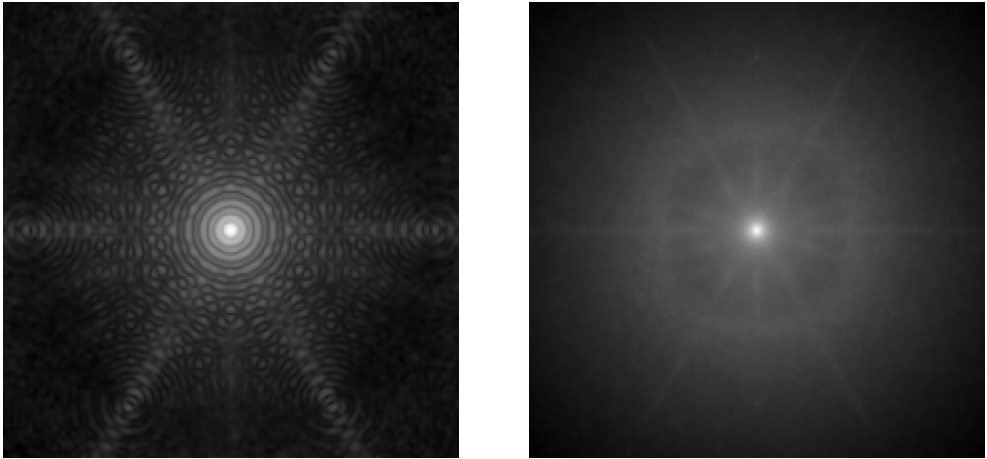}%
\caption{The IRIS PSFs used for simulations in this paper. These are the predicted PSF from the AO system at 2.2 $\micron$ (\kband, left) and 0.88 $\micron$ (\zband, right) at zenith with a Strehl ratio of 0.72 and 0.19, respectively. The images are at the 4 mas plate scale and 1$\arcsec$ in radius and are displayed on a log scale to highlight the wings of the PSF.}
\label{fig:psf}
\end{figure*}

The PSFs from the TMT NFIRAOS system are simulated at telescope zenith and expected median seeing conditions. The closed loop PSFs are first computed using standard fast-Fourier transform (FFT) methods with the Multithreaded Adaptive Optics Simulator (MAOS\footnote{See \url{github.com/lianqiw/maos}.}). The simulations are run with low order, NGS controlled modes ideally corrected. We then run sky coverage simulations to obtain the wavefront error in low order, NGS controlled modes as a function of sky coverage \citep{Wang:09}. For a given sky coverage level, the final PSF is obtained by multiplying the optical transfer function (OTF) due to the three effects : 1) high order effects obtained from MAOS simulations with ideal correction on low order modes, 2) OTF from low order wavefront errors, 3) OTFs due to uncertainties from the optical system and instrument, which are either simulated or as allocated by design requirements of TMT, NFIRAOS, and IRIS. These PSFs are then sampled onto detector taking into account the detector pixel size and charge diffusion. Examples of the resulting \kbandadj and \zbandadj PSFs used for the IRIS simulator are shown in Figure \ref{fig:psf}.

\begin{deluxetable*}{lccccccc}
\tablecolumns{8}
\tablecaption{IRIS Filters and Sky Background}
\tabletypesize{\scriptsize}  
\tablehead{\colhead{Filter} & \colhead{Integrated Background} & \colhead{Background between OH\tablenotemark{a}}  & \colhead{Zero Point\tablenotemark{b}} & \colhead{$\lambda_\mathrm{central}$}  & \colhead{$\lambda_\mathrm{cut-on}$}  & \colhead{$\lambda_\mathrm{cut-off}$}  & \colhead{Throughput\tablenotemark{c}} \\
\colhead{} & \colhead{(mag arcsec$^{-2}$)} & \colhead{(mag arcsec$^{-2}$)} & \colhead{(photons s$^{-1}$ m$^{-2}$)} & \colhead{($\micron$)} & \colhead{(\micron)} & \colhead{(\micron)} & \colhead{}}
\startdata
$Z$ & 18.8 & 18.9 & $1.37\times10^{10}$ & 0.928 & 0.840 & 1.026 & 0.351 \\
$Y$ & 17.2 & 18.5 & $9.85\times10^9$ & 1.090 & 0.980 & 1.200  & 0.351 \\
$J$ & 16.5 & 18.1 & $6.43\times10^9$ & 1.310 & 1.180 & 1.440 & 0.373 \\
$H$ & 13.7 & 16.7 & $3.97\times10^9$ & 1.635 & 1.470 & 1.800 & 0.432 \\
$K$ & 13.9 & 13.9 & $1.72\times10^9$  & 2.182 & 1.975 & 2.412 & 0.421 \\
\enddata
\tablenotetext{a}{Predicted background between OH lines.}
\tablenotetext{b}{Flux corresponding to a zeroth magnitude star in the Vega magnitude system.}
\tablenotetext{c}{Total spectroscopic instrumental throughput (including telescope, optics, detector, etc.) using the Zemax optical design. These values are from the IRIS Design Requirements Document.}
\label{tab:filters}
\end{deluxetable*}

\subsection{SNR and noise sources}

The SNR is computed using the standard approximation: 
\begin{equation}
\frac{S}{N} = \frac{S\sqrt{T}}{\left[S+\sum\limits^{n}_{i=1}(B + D + R^2/t)\right]^{1/2}}
\end{equation}
where $R$ is the readout noise ( 2 electrons), $S$ is the total signal from the object summed over all pixels (electrons s$^{-1}$), $B$ is the sky background per pixel (electrons s$^{-1}$), $D$ is the dark current per pixel (0.002 electrons s$^{-1}$), $t$ is the exposure time per frame, $n$ is the number of pixels, and $T$ is the total exposure time ($T = tN_{\mathrm{frames}}$, if each frame has the same exposure time). $B$ includes all background components detailed in Section \ref{sec:background}. For these simulations we use the predicted total throughput of the telescope, AO system, and IRIS instrument (including detector) to determine the flux observed by the detector. SNR values presented in this paper are the SNR per spaxel per wavelength channel. In order to approximate the typical spectra that will be observed in the various modes of IRIS, we use synthetic stellar spectra from the the MARCS\footnote{Downloaded from http://marcs.astro.uu.se/} stellar atmosphere models \citep{2008A&A...486..951G}. We choose to use stellar atmosphere models in order to obtain spectra with $R$ = 4000 and $R = 8000$ over the wavelength range available for observations with IRIS (0.840 $\micron$ to 2.412 $\micron$). Existing empirical spectral libraries typically have spectral resolution of 3000-4000, with higher spectral resolution available only at certain wavelength ranges.

\subsection{Background Simulation}
\label{sec:background}
We simulate the expected background on Mauna Kea using a combination of predicted and empirical measurements for the different background components:
\begin{itemize}
\item Sky background - we use the theoretical sky background available from the Gemini Observatory\footnote{More information at: \url{http://www.gemini.edu/?q=node/10787}}, which includes the sky transmission calculated from with the atmospheric model ATRAN \citep{1992nstc.rept.....L}, a 273 K continuum to simulate the sky, and Zodiacal light contribution, scaled to 18 mag arcsec$^{-2}$ at $H$ band. We use the sky background corresponding to 1.6 mm of water vapor column at an airmass of 1.5. 
\item AO system - the thermal contribution from the telescope and AO system is modeled with a modified blackbody of 275 K  ($\epsilon = 0.09$) and an AO system of 243 K ($\epsilon = 0.01$), respectively.
\end{itemize}
We compute both the integrated background values as well as the background between OH lines for the broadband filters. The integrated background is computed by integrating the total background spectrum through the filter transmission curve. While there is considerable variance in the integrated background values on Mauna Kea, the values computed here are within the observed range of backgrounds tabulated by \citet{2008PASP..120.1244S} and from the WFCAM\footnote{WFCAM sky measurements: \url{http://casu.ast.cam.ac.uk/surveys-projects/wfcam}} instrument on UKIRT. Because OH lines can dominate integrated background values, we also compute the background between the OH lines by taking the median flux of the background spectrum and using this value to integrate across the filter. These values will be useful for spectroscopic observations that can be made between OH lines. We note that the background between OH lines has been very difficult to measure with spectrographs from the ground because the flux from OH lines scatter internally in the instrument into wavelengths far from their peaks \citep{Woods:94}. For example, \citet{2008MNRAS.386...47E} showed that measurements of the continuum between OH lines in $H$ band can differ by more than a order of magnitude between the ground and space. The background values for all broadband filters are tabulated in Table \ref{tab:filters}.

\subsection{Galaxy simulations}

The primary goal of these simulations is to reproduce the nuclei of galaxies as they would be observed with IRIS. As these cores are well resolved with IRIS, the SNR is a strong function of the surface brightness of their inner nuclei. Because IRIS has much smaller pixel scales than typically used for present day galaxy observations, depending on how concentrated the light distribution is in their cores, the amount of signal per spaxel can be lower than a corresponding spaxel (typically between 50-200 mas) in IFSs in use today. In order to determine the surface brightness at the angular scales of IRIS observations, we extrapolate the existing published empirical fits to light profiles of galactic nuclei of interest to the inner region that would be sampled by IRIS. We use surface brightness profiles measured in \kband whenever possible, but for galaxy samples such as those compiled by \citet{2007ApJ...664..226L} in $V$ band, we use a color of $V - K = 3$ to convert into \kbandadj surface brightnesses (see Section \ref{sec:nearby_galaxies}). This color corresponds approximately to a K3III late-type giant, typical of the spectra observed for early-type galaxies.

\begin{figure}[!ht]
\centering
\includegraphics[width=3.4in]{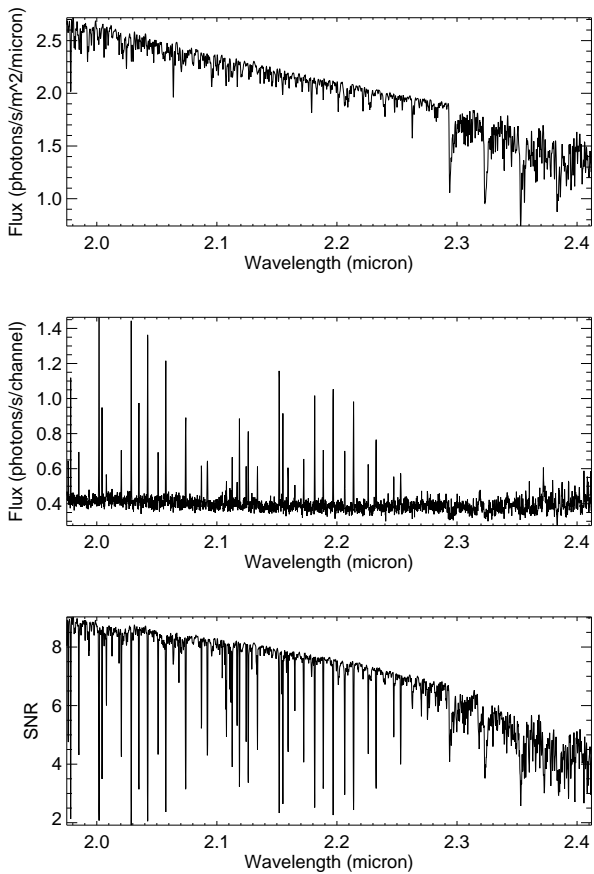}
\caption{Top: synthetic spectrum of a $\teff = 4500 K$ ($\sim$K3III) giant used for the simulations. The fluxes are given for observations of a location with surface brightness with K = 13 mag arcsec$^{-2}$,  at a spectral resolution $R = 8000$. Middle: The spectrum of the background and noise sources (see Sections \ref{sec:background}) simulated for an observation with a plates scale of 9 mas with an integration time of 900 s. Bottom: the resulting SNR at each spectral channel in a single spaxel. The sharp dips in the SNR are due to the higher sky background at the wavelengths of the OH lines.}
\label{fig:sample_snr}
\end{figure}

\begin{figure}[!th]
\centering
\includegraphics[angle=90,width=3.4in]{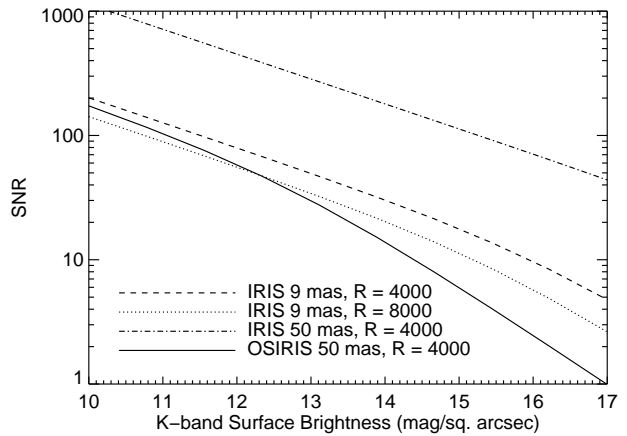}
\caption{The average SNR per spaxel per average spectral channel for extended sources with uniform surface brightness with 5 hours of total integration time (20 observations of 900 s) for an M0III stellar template source for different IRIS spaxel scales and spectral resolutions: 9 mas, $R = 4000$ (solid); 9 mas, $R$ = 8000 (dotted); 50 mas, $R$ = 4000 (dot-dashed). For comparison, we also plot the extended source sensitivity of the OSIRIS IFS on Keck (dashed).}
\label{fig:snr_extended}
\end{figure}

\begin{figure}[!th]
\centering
\includegraphics[angle=90,width=3.4in]{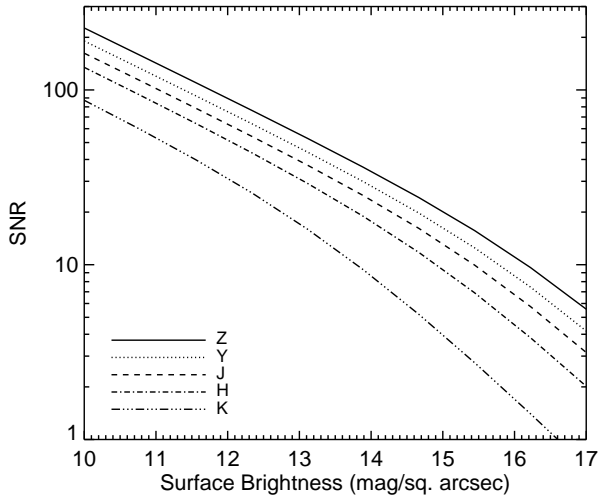}
\caption{The average SNR per spaxel per average spectral channel for extended sources with uniform surface brightness using the Z (solid), Y (dotted), J (dashed), H (dot-dashed), and K (dot-dot-dot-dashed) filters. These curves are calculated for the 4 mas plate-scale at a spectral resolution of 4000 and $20\times900$ s integrations (5 hours).}
\label{fig:snr_extended_compare}
\end{figure}

In Figure \ref{fig:sample_snr}, we show an example of a simulated 900 s observation of an extended source with \kbandadj surface brightness of 13 mag arcsec$^{-2}$ with a K3III stellar spectrum. For a single observation of 900 s with spaxel of 9 mas and a spectral resolution $R = 8000$ (3200 spectral channels between $1.975-2.412$ \micron), the average SNR per spectral channel per spaxel is $\sim6$. We show in Figure \ref{fig:snr_extended} the extended source sensitivity as a function of \kbandadj surface brightness for the 9 mas and 50 mas spaxel scales as well as $R = 4000$, and $R = 8000$. For comparison, we also plot the OSIRIS 50 mas, $R = 4000$ surface brightness sensitivity (these were calculated using values from the OSIRIS Manual, with read noise of 10 electrons, dark current of 0.31 electrons s$^{-1}$, a total throughput of 8.8\%, and a telescope collecting area of 76 m$^2$ for Keck). We find that IRIS at the 9 mas plate scale will have comparable sensitivity to OSIRIS at the 50 mas plate scale. IRIS is able to achieve similar sensitivity at 5 times better spatial sampling because of a combination of a larger telescope aperture, higher throughput, and lower detector noise. For comparison, we also compute the SNR for extended sources with uniform surface brightness for all five broad-band filters on IRIS (for the 4 mas plate-scale, $R = 4000$, $20\times900$ s integration time; Figure \ref{fig:snr_extended_compare}). 

\begin{figure}[!th]
\centering
\includegraphics[angle=90,width=3.4in]{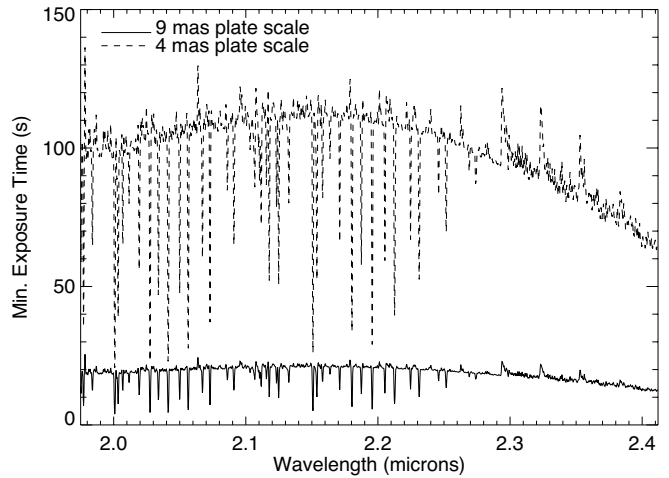}
\caption{The minimum integration time, $t_{int}$ necessary to be background limited, with background signal greater than the combination of dark current and the read noise ($B > D + R^2/t_{int}$) for the 4 mas (dashed) and 9 mas (solid) plate scale at $R = 8000$ in the \kband. The sharp dips are from the OH sky lines, which increases the amount of sky background at those wavelengths.}
\label{fig:min_tint}
\end{figure}

\subsection{Gauss-Hermite moments and spectral resolution}
\label{sec:gauss_hermite}
\begin{figure*}[!ht]
\centering
\includegraphics[angle=90,width=6in]{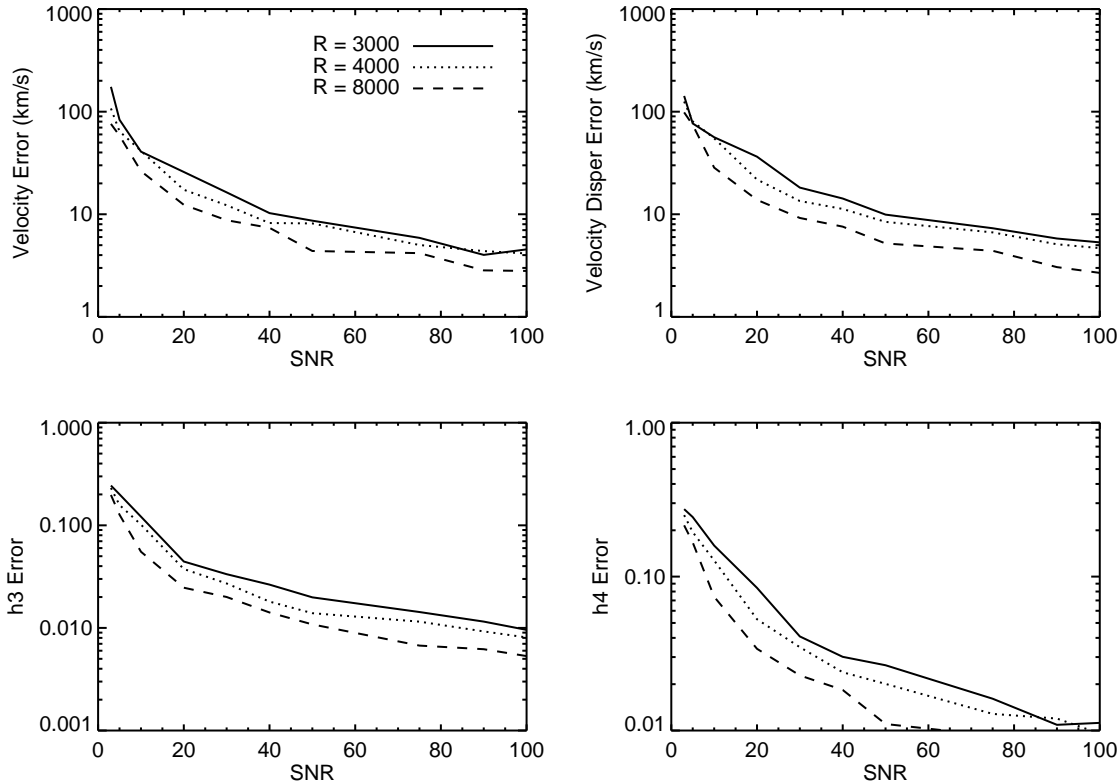}
\caption{Results of simulations of the dependence of the uncertainties in measured velocity moments on the SNR of the observed spectra at spectral resolutions of $R = 3000$ (solid), $R = 4000$ (dotted), and $R = 8000$ (dashed). The simulations were run with Gauss-Hermite moments: $v$ = 0 km s$^{-1}$, $\sigma = 350$ km s$^{-1}$, $h_3 = -0.14$, and $h_4 = 0.03$. These would be for the most massive galaxies. Uncertainties in velocity dispersion drop below 10 km s$^{-1}$ with a SNR $>$ 30 for $R = 8000$ (top right). }
\label{fig:disp_err_sim_350}
\end{figure*}

\begin{figure*}[!ht]
\centering
\includegraphics[angle=90,width=6in]{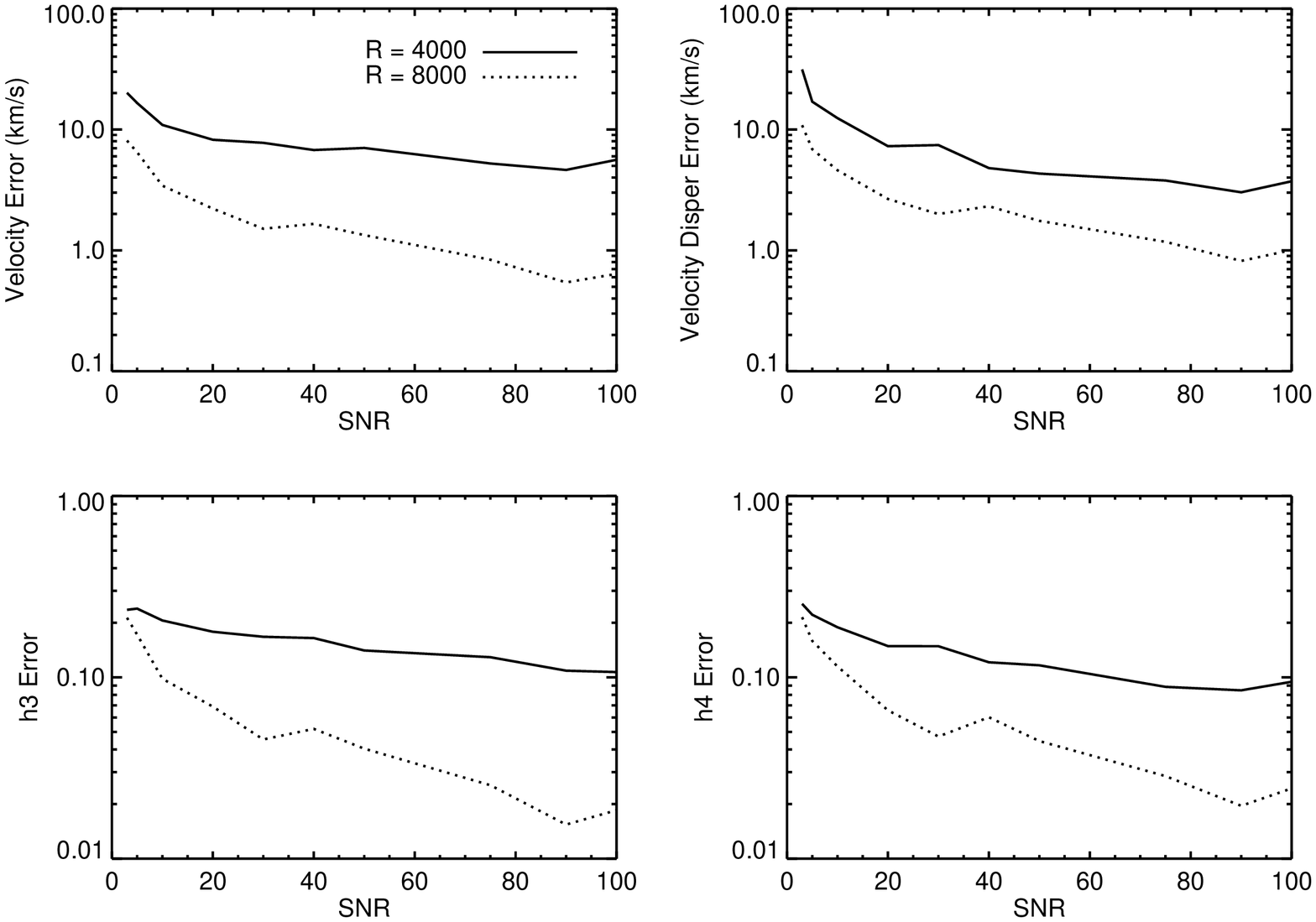}
\caption{Results of simulations of the dependence of the uncertainties in measured velocity moments on the SNR of the observed spectra at spectral resolutions of $R = 8000$ (solid) and $R = 4000$ (dotted). The simulations were run with Gauss-Hermite moments: $v$ = 0 km s$^{-1}$, $\sigma = 30$ km s$^{-1}$, $h_3 = -0.14$, and $h_4 = 0.03$. Uncertainties in velocity dispersion drop below 2 km s$^{-1}$ with a SNR $>$ 30 for $R = 8000$ (top right). Because of limited spectral resolution, the $R = 4000$ simulation was not able to obtain robust measurements of $h_3$ or $h_4$ even at high SNR. Dynamical measurements of intermediate-mass black holes or low mass nuclei will require the $R = 8000$ mode to be effective.}
\label{fig:disp_err_sim}
\end{figure*}

The ability of stellar kinematic measurements from spectroscopy to derive black hole masses rests largely on how well the line-of-sight velocity distribution (LOSVD) can be determined \citep[e.g.][]{1993ApJ...407..525V,2000AJ....119.1157G}. The shape of the LOSVD is a signature of the underlying orbits of unresolved stars under the gravitational influence of the central black hole the extended mass distribution of stars, gas, and dark matter. The LOSVD also holds information about the velocity anisotropy of the system, which can be degenerate with the black hole mass when using velocity dispersions alone to constrain the potential. Typically, the LOSVD at each spatial location is either derived non-parametrically by fitting to orbit libraries  \citep{2000AJ....119.1157G} or it is measured from the spectra by describing the LOSVD as Gauss-Hermite moments \citep{1993ApJ...407..525V}. Both methods lead to comparable precision in the measurements of the black hole mass \citep[e.g.][]{2011MNRAS.410.1223R,2011Natur.480..215M}. In order to investigate how the SNR and spectral resolution of IRIS affect the determination of LOSVD, we use the latter method to simulate different LOSVD with the first 4 Gauss-Hermite polynomials: the mean velocity $v$, the velocity dispersion $\sigma$, $h_3$, and $h_4$. The LOSVD is defined as follows by \citet{1993ApJ...407..525V}:
\begin{equation}
\mathcal{L}(v) = \frac{e^{-y^{2}/2}}{\sigma \sqrt{2\pi}}[1 + \sum_{m=3}^M h_m H_m(y)],
\end{equation}
where $ y = (v - V)/\sigma$, and $H_m(y)$ are the Hermite polynomials defined as:
\begin{eqnarray}
& H_0(y) = 1 & \\
& H_1(y) = \sqrt{2}y & \\
& H_2(y) = \frac{1}{\sqrt{2}} (2y^2-1) & \\
& H_3(y) = \frac{1}{\sqrt{6}}(2\sqrt{2} y^3 - 3\sqrt{2} y) & \\
& H_4(y) = \frac{1}{\sqrt{24}}(4 y^4 - 12 y^2 + 3) &
\end{eqnarray}
The velocity dispersion along the line of sight results from the projected motions of stars under in a gravitational potential. The parameters $h_3$ and $h_4$ measures the deviation from a Gaussian distribution of velocities: $h_3$ measures the skew in the distribution and is related to the amount of anisotropy, while $h_4$ measures the excess in the number of stars with high velocities in the wings of the Gaussian.  We investigate the precision that IRIS would be able to measure these Gauss-Hermite moments as a function of spectral resolution and SNR. This is accomplished by convolving a synthetic spectrum of a 4500 K giant with a LOSVD, adding random noise, and recovering the Gauss-Hermite moments. We fit for the Gauss-Hermite moments using the Penalized Pixel-Fitting code from \citet{2004PASP..116..138C}. The code works by iteratively fitting a combination of template spectra convolved with different velocity moments and noise to obtain the best fit kinematic parameters. We input a series of synthetic spectra with temperatures ranging from 3100-4500 K in order to allow the algorithm freedom in choosing the best fit template as well as to simulate possible template mismatches. We find that for observations of nuclei with velocity dispersion of 200-350 km s$^{-1}$, which corresponds to $> 10^7$ M$_{\odot}$, $R = 4000$ should be sufficient to obtain uncertainties in $\sigma \sim 5-6$ km s$^{-1}$, and $h_3$ and $h_4$ of $\sim0.02$ at SNR $> 60$, comparable to that of IFS measurements today at the same spectral resolution (for simulated $v = 0$ km s$^{-1}$, $\sigma = 350$ km s$^{-1}$, $h_3 = -0.14$, $h_4 = 0.03$, see Figure \ref{fig:disp_err_sim_350}). For the low velocity dispersion case ($v = 0$ km s$^{-1}$, $\sigma = 30$ km s$^{-1}$, $h_3 = -0.14$, $h_4 = 0.03$), appropriate for intermediate mass black holes and nuclear star clusters, we find that a spectral resolution of $R = 4000$ will be insufficient to accurately determine $h_3$ and $h_4$ at any SNR. These observations will require the higher spectral resolution mode of $R=8000$, which will able to achieve uncertainties in velocity dispersion, $\Delta\sigma \sim 1$ km s$^{-1}$ for SNR $\gtrsim 50$ and uncertainties in $h_3$ and $h_4$ of less than 0.03 (see Figure \ref{fig:disp_err_sim}). This spectral resolution is higher than any of the current IFSs and should enable IRIS to expand the sample of observable galaxies to those with low velocity dispersions.

\section{Results}
\label{sec:results}
\subsection{Intermediate Mass Black Holes ($< 10^6$ \msun)}

Intermediate mass black holes have been a source of debate in recent years, because while they have been predicted  to exist theoretically, the observational evidence for IMBHs is subject to much debate \citep[e.g.][]{2010ApJ...710.1032A}. These objects are interesting because they can tell us about the formation mechanisms of massive black holes \citep{2010A&ARv..18..279V}. We know that supermassive black holes and stellar mass black holes must have been created through different physical mechanisms because of their great difference in mass. IMBHs on the other hand are of $10^3 - 10^5$ \msun, occupying an interesting position in between the extreme mass ranges of stellar and supermassive black holes. Theoretically, they should exist in globular clusters or small compact galaxies if we extrapolate from the M-$\sigma$ relation measured using supermassive black holes. 

\begin{figure}[!th]
\centering
\includegraphics[angle=90,width=3.4in]{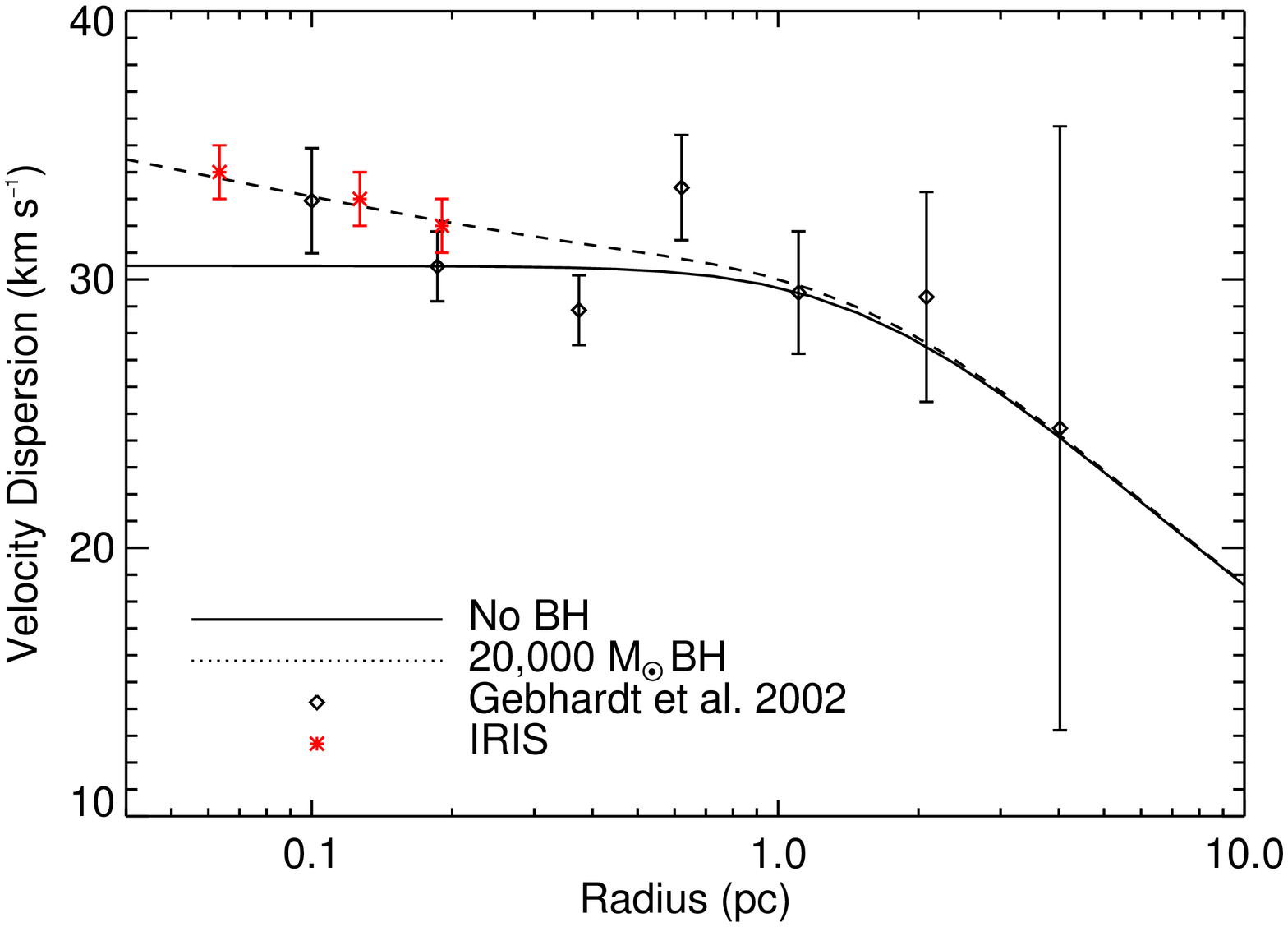}
\caption{The measured dispersion profile from \citet{2002ApJ...578L..41G}, as reproduced by \citet{2003ApJ...589L..25B}, is plotted as diamond points on top of the expected velocity dispersion curve for a star cluster with a black hole (dashed) and one without. The star cluster is modeled with a King profile with isotropic velocity dispersion and a total mass of $\sim 7.6\times10^6$ \msun \citep{2003ApJ...589L..25B}. The previous measurements are only marginally within the region of Keplerian falloff in velocity dispersion profile. Expected IRIS \kbandadj measurements are shown in red asterisks, which will be able to sample significantly further into the potential providing the critical measurements necessary to measure the presence of an IMBH.}
\label{fig:g1_disp}
\end{figure}

G1, likely the most massive globular cluster in M31 \citep[$\sim 7.6\times10^6$ \msun][]{2003ApJ...589L..25B}, has been a source of great interest because it may potentially harbor an IMBH at its center \citep{2002ApJ...578L..41G}. At 770 kpc, it is also one of the closest targets for dynamical searches for IMBHs. \citet{2002ApJ...578L..41G} first reported that based on spectroscopy from STIS on \textit{HST}, the cluster contains an IMBH with a mass of about $2\times10^4$ \msun. However, a different dynamical modeling by \citet{2003ApJ...589L..25B} found that the kinematic measurements can be consistent with a stellar cluster alone. The difficulty in determining whether an IMBH exists in G1 stems from the low angular resolution of the observations compared to the gravitational radius of influence of a $10^4$ \msun black hole. Figure \ref{fig:g1_disp} shows the velocity dispersion measurements from \citet{2002ApJ...578L..41G}, as reproduced by \citet{2003ApJ...589L..25B}. The observed radial profile of the velocity dispersion is plotted on top of two model dispersion profiles - one for a star cluster with a King profile with a black hole and one without. Only the last point in the measured dispersion profile deviates slightly above that predicted from the star cluster alone. 

\begin{figure*}[!tbp]
\centering
\includegraphics[angle=90,width=6.5in]{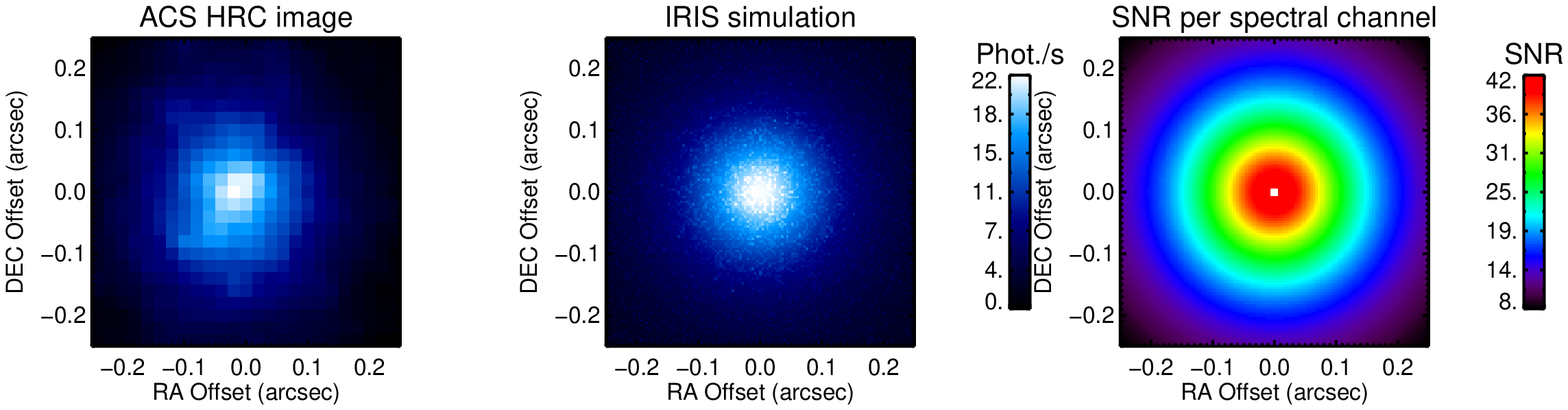}
\caption{\textbf{Left}: an ACS image of the central 0.$\arcsec$25 of the G1 globular cluster (0.9 pc). \textbf{Center}: Simulated image of the same region of the cluster with 4 mas pixels and based on the best fit King profile to the \kbandadj surface brightness of G1 extrapolated toward the center. Poisson noise was added to this image based on the expected source counts as well as the background counts. The simulation uses the R = 8000 mode, and 10 exposures with 900 s integrations each ($\sim$2.5 hr total). \textbf{Right}: the expected SNR per spaxel per spectral channel in the continuum. This SNR should translate to a velocity dispersion uncertainty $< 2$ \kms (see Section \ref{sec:gauss_hermite}).}
\label{fig:g1_surf}
\end{figure*}

In order to assess the capabilities of IRIS and TMT for observing this cluster, we simulate G1 using the observed \kbandadj luminosity profile and a simple dynamical model for an isotropic cluster with a King surface density profile \citep{1962AJ.....67..471K}:
\begin{equation}
I_K(R) = I_o \left [\frac{1}{(1+(\frac{R}{R_c})^2)^{\frac{1}{2}}} - \frac{1}{(1+(\frac{R_t}{R_c})^2)^{\frac{1}{2}}} \right ],
\end{equation}
where $R$ is the projected distance from the cluster core, $R_c$ is the core radius, $R_t$ is the tidal radius, and $I_o$ is the normalization to the flux. The deprojected volume density profile is:
\begin{eqnarray}
& n(r) = \frac{n_o}{\pi r_c [1+(r_t/r_c)^2]^{\frac{3}{2}}}\frac{1}{z^2}\left [ \frac{1}{z}\cos^{-1}z - (1-z^2)^{\frac{1}{2}}\right ], & \\ 
& z = \left[ \frac{1+(r/r_c)^2}{1+(r_t/r_c)^2}\right]^{\frac{1}{2}}, &
\end{eqnarray}
where $r$ is the physical radius. We use the \kbandadj surface brightness from the 
\textit{HST}/NICMOS observations of \citet{2001AJ....121.2597S} to infer a surface brightness of K = 11 mag arcsec$^{-2}$ at the core radius in order to normalize the surface brightness profile for the SNR calculation.  We use the cluster parameters from the profile fits of \citet{2001AJ....122..830M} and the cluster mass of $8\times10^6$ \msun from \citet{2003ApJ...589L..25B} to model the velocity dispersion profile for a cluster with and without a central black hole of $2\times10^4$ \msun. The projected line of sight velocity dispersion is calculated from the Jeans equation for an spherically symmetric and isotropic cluster \citep[see e.g.][]{2009A&A...502...91S}:
\begin{equation}
\sigma_p^2(R) = \frac{2G}{\Sigma(R)}\int_R^\infty \frac{ r dr}{\sqrt{r^2-R^2}}\int_r^\infty \frac{dr^\prime M(r^\prime) n(r^\prime)}{r^{\prime 2}},
\end{equation}
where $\Sigma(R)$ is the projected surface density of the tracer stars (equivalent to the King profile in this case), and $M(r)$ and $n(r)$ are the spatial mass and density profile, respectively. The actual measurement with IRIS will involve a much more sophisticated dynamical model of the cluster, but this model serves as a good illustration of the power of higher angular resolution measurements in resolving the sphere of influence of intermediate mass black holes. We find that IRIS on TMT will have the angular resolution to observe the Keplerian fall off in the velocity dispersion profile due to the mass of a simulated IMBH (see Figure \ref{fig:g1_disp}). With the 4 mas plate scale, $R = 8000$ mode, and 10 exposures with 15 min integrations each ($\sim 2.5$ hr total), IRIS will be able to obtain about 1-2 km s$^{-1}$ velocity dispersion uncertainties at these critical points, providing a definitive measurement of the black hole mass if one is present in G1. Figure \ref{fig:g1_surf} shows the comparison between an image of G1 taken using \textit{HST}/ACS and a simulated image with IRIS showing the increase in angular resolution that will be provided by TMT. 

\subsection{Milky Way-like black holes ($10^6-10^7$ \msun)}

Presently, there are only a few galaxies with dynamical black hole mass measurements in the range of $10^6 - 10^7$ \msun. Because these galaxies lie at the lower mass end of the M-$\sigma$ relationship, they have a large impact on the slope of the relationship. The Milky Way, while it has a precise black hole mass measurement, has a highly uncertain bulge velocity dispersion measurement given that we are situated within the Galaxy \citep[e.g.][]{2002ApJ...574..740T}. A larger sample of black holes of a comparable mass range would greatly improve not only the precision of the measured M-$\sigma$ relationship, but also our understanding of black hole demographics in this mass range. 

\begin{figure*}[th]
\centering
\includegraphics[angle=90,width=6.5in]{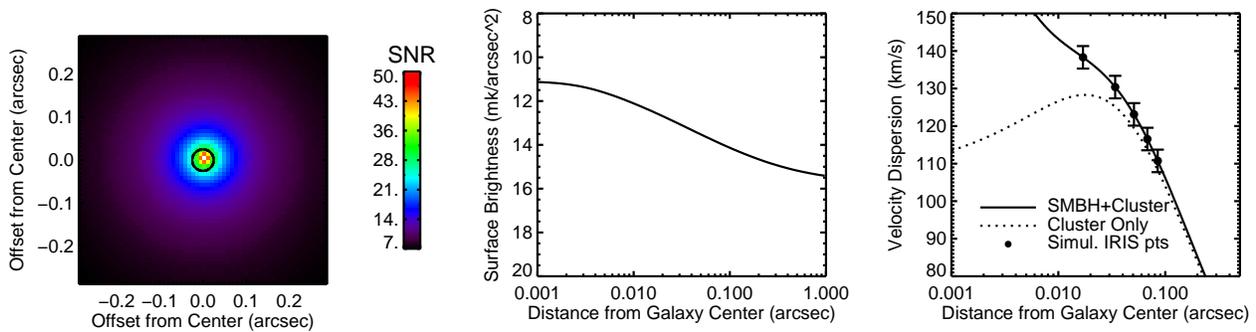}
\caption{\textbf{Left}: Simulation of the SNR of observations of the extinction corrected core of the Milky Way if it were at the distance to the Virgo Cluster ($\sim16$ Mpc), at \kband with 10 observations of 900 s each, at $R = 4000$ in the 9 mas plate scale. The black circle is the marks the radius of influence of a $4\times10^6$ \msun BH. \textbf{Center}: \kbandadj surface brightness of the Milky Way core at the distance to Virgo. \textbf{Right}: the velocity dispersion curve of the nuclear star cluster with a $4\times10^6$ \msun black hole and a star cluster with $\sim1\times10^6$ \msun within the central 1 pc (solid line). The dashed line shows the velocity dispersion profile for the cluster alone. The innermost measurements from IRIS will be able to sample the Keplerian falloff the in velocity dispersion profile.}
\label{fig:mw}
\end{figure*}

Our simulation of the Milky Way nuclear star cluster uses the \kbandadj surface brightness profile measured by \citet{1999A&A...348..768P} and \citet{2010arXiv1001.4238S}. In order to simulate the velocity dispersion measurements, we use the isotropic cluster model from \citet{2009A&A...502...91S} and their parameterization of the spatial density profile in a broken power law model:
\begin{equation}
n(r) = n_o \left(\frac{r}{r_o}\right)^{-\gamma}\left(1+ \frac{r}{r_o}\right)^{\gamma-A}, \\
\end{equation}
with A = 1.8, $\gamma$ = 0.5, and $r_o$ = 20$\arcsec$ (0.8 pc). We also correct for the average extinction of A$_K$ = 2.7 \citep{2009A&A...502...91S}, to simulate the case of a Milky Way-like galaxy that would be more likely to be observed in a more optimal inclination than the edge-one view we have of the Galactic center. We find that IRIS would be able to detect the Keplerian fall-off in the velocity dispersion using the 4 mas pixel scale at sufficient SNR to observe an $~\sim 4\times10^6$ \msun black hole at the distance to the Virgo Cluster (16.5 Mpc), if it contains a star cluster as compact as one at the center of our Galaxy (with $\sim 2.5$ hr of integration time, see Figure \ref{fig:mw}). 

\begin{deluxetable*}{lcccccccccccc}
\tablecolumns{13}
\tablecaption{SNR simulations for nearby bright early-type galaxies}
\tablewidth{0pc}
\setlength{\tabcolsep}{0.01in} 
\tabletypesize{\scriptsize}  
\tablehead{\colhead{Name\tablenotemark{a}} & \colhead{Type} & \colhead{M$_V$} & \colhead{I$_{b,K}$} & \colhead{Distance} & \colhead{Scale\tablenotemark{b}} & \colhead{$\gamma$\tablenotemark{c}} & \colhead{ Peak SNR\tablenotemark{d}} & \colhead{Int. SNR\tablenotemark{e}} & \colhead{Peak SNR\tablenotemark{f}} & \colhead{Int. SNR\tablenotemark{g}} & \colhead{Pred. M$_{BH}$\tablenotemark{h}} & \colhead{$r_{infl}/res.$\tablenotemark{i}} \\
\colhead{} & \colhead{} & \colhead{} & \colhead{} & \colhead{(Mpc)} & \colhead{(pc/res.)} & \colhead{} & \colhead{4 mas} & \colhead{4 mas} & \colhead{9 mas} & \colhead{9 mas} & \colhead{(\msun)} & \colhead{}}
\startdata
\input{lauer_galaxy_sample_table_example}
\enddata
\tablenotetext{a}{Galaxy sample from \citet{2007ApJ...664..226L}, along with galaxy properties (Type, bulge M$_V$, Distance, and $\gamma$). Observations are simulated using 8 observations of 900 s each for a total integration time of 2 hr. The full table is provided electronically.}
\tablenotetext{b}{Physical scale of one resolution element (18 mas at \kband) at the distance to the galaxy.}
\tablenotetext{c}{Inner power law slope of best fit Nuker profile \citep{2007ApJ...664..226L}.}
\tablenotetext{d}{SNR of the peak spaxel in the sensitivity simulations using the 4 mas plate scale.}
\tablenotetext{e}{SNR from all spaxels within a resolution element in the sensitivity simulations using the 4 mas pixel scale.}
\tablenotetext{f}{SNR of the peak spaxel in the sensitivity simulations using the 9 mas plate scale.}
\tablenotetext{g}{SNR from all spaxels within a resolution element in the sensitivity simulations using the 9 mas pixel scale.}
\tablenotetext{h}{Predicted black hole mass based on the M$_{BH}$--L relationship, $\log{(M_{bh}/M_\odot)} = 8.98 + 1.11 \log{(L_V/10^{11} L_{\odot,V})}$ from \citet{2009ApJ...698..198G}.}
\tablenotetext{i}{The number or resolution elements within the radius of influence of the black hole at \kband ($\sim18$ mas).}
\label{tab:nearby_early_type}
\end{deluxetable*}

\subsubsection{Nuclear Star Clusters}

\begin{figure*}[th]
\centering
\includegraphics[angle=90,width=5.5in]{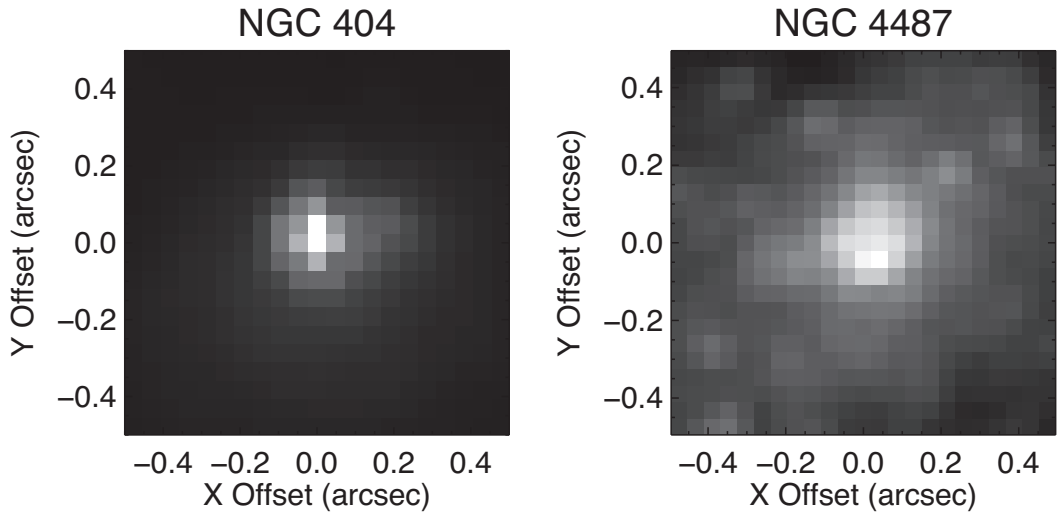}
\caption{Examples of nearby nuclear star clusters. \textbf{Left:} ACS F658N image of the nucleus of NGC 404, with a distance of 3.06 Mpc (image width: 15 pc). \textbf{Right:} WFPC2 F814W image of the nucleus of NGC 4487 at a distance of 14.6 Mpc (image width: 71 pc). With half-light radii less than 5 pc, the compactness of nuclear star clusters necessitate high angular resolution observations to model their kinematics.}
\label{fig:nsc_example}
\end{figure*}

The extrapolation of the black hole masses to less luminous galaxies and smaller bulges would predict less massive central black holes in these galaxies. Instead, many of these galaxies are found to have compact star clusters at their centers, with half-light radii of 3-10 pc with masses of $10^6-10^7$ \msun \citep[e.g.][]{2004AJ....127..105B}. These clusters are rather massive for star clusters and intriguingly, their masses appear to follow the \msigma relationship, leading to the proposal that they are a product of the same galaxy evolution mechanism that produces supermassive black holes and their scaling relationships with galaxy properties \citep{2006ApJ...644L..21F,2006ApJ...644L..17W}. Understanding the origins of these clusters would then help us understand the coupling of the nuclei to galaxy evolution. However, a number of questions remain about the basic properties of these clusters; for example, it is unclear at this time whether these low mass galaxies with nuclear star clusters have central black holes as well. Locally, we know that M33 likely does not contain a central black hole \citep[upper limit \mbh $< 1500 \msun$,][]{2001AJ....122.2469G,2001Sci...293.1116M}, while M32 has both a nuclear star cluster and a massive black hole of $\sim2.5\times10^6$ \msun \citep[e.g.][]{2002MNRAS.335..517V,2010MNRAS.401.1770V}. These clusters represent a major observational challenge since they are faint, compact, and any likely black holes will have relatively low masses. Resolving the sphere of influence of these black holes is generally out of reach with current telescopes (see Figure \ref{fig:nsc_example}). Only a few sources in the local universe have constraining limits on their black hole masses \citep[e.g.][]{2009MNRAS.397.2148G,2010ApJ...714..713S}. TMT will provide both the sensitivity and angular resolution to observe some of these low mass nuclei that are unaccessible today. A spectral resolution of $R = 8000$ will likely also be required to measure accurate kinematic moments. 

\subsection{Massive black holes ($>10^7$ \msun)}
\label{sec:nearby_galaxies}

In order to investigate the ability for IRIS to detect black holes in nearby early-type galaxies, we used the sample of 219 early-type galaxies compiled by \citet{2007ApJ...664..226L}. This sample includes surface brightness fits for S0 and elliptical galaxies using images from WFPC2, WFPC1, and NICMOS on \textit{HST}. We use their best fit Nuker profile in order to extrapolate these data to the center of the galaxies at the resolution of TMT. The Nuker profile is defined as: 
\begin{equation}
I(r) = I_b 2^{(\beta - \gamma)/\alpha} \left(\frac{r}{r_b}\right)^{-\gamma} \left[1 + \left(\frac{r}{r_b}\right)^\alpha \right]^{(\gamma - \beta)/\alpha},
\end{equation}
where $\gamma$ is the inner slope, $\beta$ is the outer slope, $\alpha$ controls the strength of the abruptness of the transition between the two slopes, $r_b$ is the break radius, and $I_b$ is the intensity at the break radius. The parameter most relevant for these simulations is $\gamma$, which controls the steepness of the light profile at the center of the galaxy. Since the data from \citet{2007ApJ...664..226L} are given in $V$-band magnitudes, we convert these to the $K$-band using $V - K = 3.0$, the typical color of a K-type giant. The simulations are done at \kband at both the 4 and 9 mas plate scales. We assume each galaxy is observed 8 times with an integration time of 900 s for each spectrum (total integration time of 2 hr). 

The resulting simulated data are summarized in Table \ref{tab:nearby_early_type}. We tabulate the physical scale per $K$-band resolution element given the distance from \citet{2007ApJ...664..226L} (an angular resolution of 18 mas corresponds to about 0.87 pc at a distance of 10 Mpc). We also tabulate the SNR per spectral channel (outside of OH lines) of the peak spatial-pixel at the center of the galaxies at the 4 and 9 mas plate scales. We also include the value of the integrated SNR of all spatial-pixels within the central resolution element. For many of these galaxies, the 9 mas pixel scale will provide an optimal Nyquist sampling of the PSF at \kband and provide for a high SNR measurement within the black hole's sphere of influence. 

\begin{figure*}[th]
\centering
\includegraphics[angle=90,width=6.5in]{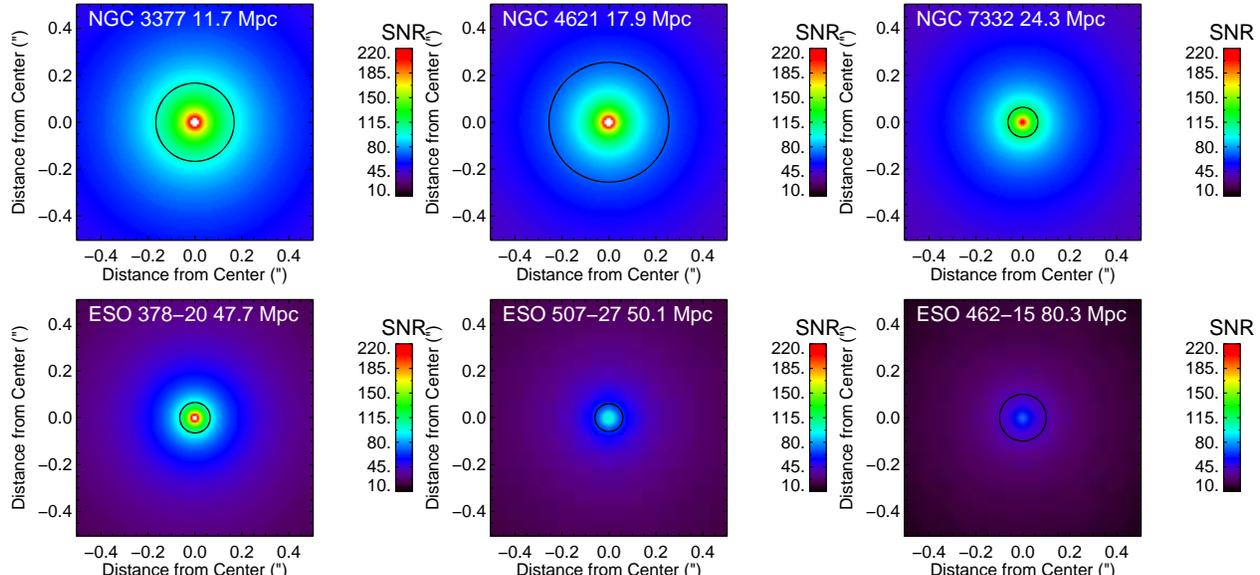}
\caption{Simulated average SNR of observations of nearby early-type galaxies with the 9 mas plate scale mode of IRIS in the \kband with 8 observations of 900 s each (2 hours integration time) at $R = 4000$. The black circle at the center represent the radius of influence of the black hole predicted based on the galaxy's luminosity. The physical scale of the width of each image is: NGC 3377 -- 57 pc, NGC 4621 -- 87 pc, NGC 7332 -- 118 pc, ESO 378-20 -- 231 pc, ESO 507-27 -- 243 pc, ESO 462-15 -- 389 pc.}
\label{fig:9mas_example}
\end{figure*}

Based on the existing measurement of the $M_V-\mbh$ relationship from \citet{2009ApJ...698..198G} ($\log{(M/M_\odot)} = 8.95 + 1.11\log{(L_V/10^{11} L_{\odot,V})}$), we convert the observed $M_V$ values into predicted central black hole masses in order to obtain a radius of influence to assess their observability. We use the luminosity because the $\mbh-L_V$ relationship appears to be a slightly better predictor of the mass of the most massive black holes \citep{2007ApJ...662..808L}. To obtain the luminosity, we use the relationship between the luminosity and the extinction corrected absolute $V$-band magnitude of the bulge from \citet{2009ApJ...698..198G}: $\log(L_v/L_{\odot,V})= 0.4(4.83 - M_{V,bulge}^0$). In order to determine the sphere of influence of the predicted black holes, we use the \msigma relationship to remove the dependence on $\sigma$ in $r_{infl} = GM/\sigma^2$. We plot the radius of influence for a sample of galaxies in Figure \ref{fig:9mas_example} and tabulated the number of resolution elements that IRIS will be able to observe within the sphere of influence in Table \ref{tab:nearby_early_type}. IRIS will be able to well resolve the sphere of influence for many of the predicted black hole masses with more than 2 resolution elements, and in some cases greater than 10 resolution elements.

\subsubsection{Brightest cluster galaxies (BCGs): an example observing program}
The dearth of measurements of black hole masses $> 10^9$ \msun is limiting our current ability to understand the origins of the black hole scaling relations and the evolution of supermassive black holes. Observations at high redshifts show that there may be black holes as massive as $10^9$ \msun at $z > 6$ \citep[e.g.,][]{2006AJ....132..117F,2010A&ARv..18..279V}. This would imply that there should exist $> 10^{10}$ \msun black holes today at the centers of the most massive galaxies. Indeed, some of the most massive black holes found today are in the massive bright galaxies ($\sim10^{11}$ L$_\odot$) at the centers of galaxy clusters, or brightest cluster galaxies (BGGs). Black holes with masses $>10^{10}$ \msun have been recently discovered in such galaxies \citep{2011Natur.480..215M,2012ApJ...756..179M}, though their frequency and formation are still unknown. Nearby in the second brightest galaxy in the Virgo cluster, M87 may contain a $6.6\pm0.4\times10^9$ \msun black hole as measured using stellar kinematics \citep{2011ApJ...729..119G}. However, gas measurements result in a factor of 2 lower mass \citep{2012ApJ...753...79W}. These galaxies are important to study as the \mlum and the \mlum relationship predict very different central black hole masses when extrapolated to galaxies at these masses. 

As example of the advantage of the increase in sensitivity and angular resolution from IRIS and TMT, we present an observing program to measure the BH masses of a statistically large sample of BCGs to investigate the uncertainties described above. We will utilize the HST survey of \citet{2003AJ....125..478L}, which presents detailed surface brightness profiles fits to 60 nearby BCGs. We convert the $I$ band surface brightness profile to \kband surface brightness using the relationship $I - K = 1.6$, typical for late-type stars. Given the range of expected BH masses ($10^9$ to $10^{10}$ \msun), and the distances to these galaxies ($< 170$ Mpc), the sphere of influence of these BHs will be spatially resolved with IRIS.  For each galaxy, we compute the integration time necessary to achieve a SNR of about 40. This SNR is based on the requirements from dynamical modeling today (Section \ref{sec:current}). We find that 51 sources will be observable with an integration time of less than 5 hours, using individual frames of 900 s each. 21 sources will require less than 1 hr of total integration time. Based on the top level requirements from the IRIS Design Requirements Document, the overhead for slewing and target acquisition with the AO system is  conservatively estimated to be about 10 min per source. The estimate for dithering and readout is about 1 min per frame. Observations of standard stars will take about 20 min per night. We also estimate that sky observations will be required at least once per source and every 2 hours. The total required observing time for 51 sources is about 139 hr, or 14 nights (to include the remaining 9 sources from \citet{2003AJ....125..478L} would require about 33 nights). Given that such a large sample can be observed with a relatively modest amounts of time, IRIS will be ideal to study the high mass end of the BH scaling relationships.

\section{Estimating the demography of black hole measurements with SDSS}
\label{sec:sdss}
While the sample of galaxies observed with \textit{HST} are well characterized at high spatial resolution, they only serve as a small sample of the possible range of galaxies that will be accessible with upcoming GSMTs. In order to determine the demography of galaxies with black hole masses measurable with stellar dynamics, we turn to the Sloan Digital Sky Survey \citep[SDSS DR7,][]{2009ApJS..182..543A}. While the spatial resolution of the SDSS galaxies is not well matched to the GSMTs, it provides the most comprehensive sample of galaxies to study statistically the number of galaxies that will be good targets for GSMTs. To extrapolate to the nuclear regions of the SDSS galaxies, we use the sample of galaxies from \citet{2011ApJS..196...11S}, who examined over 1 million SDSS DR7 galaxies and provided Sersic profile fits to the surface brightness profiles at $r$ and $g$ bands for over 600,000 galaxies. For the sources in our study, we use only the bulge component of the disk-bulge decomposition of \citet{2011ApJS..196...11S}, as it is the region where the nucleus is embedded.  We exclude galaxies with $z < 0.005$ because the peculiar motion of the galaxy begins to dominate over cosmological redshift making the distance determination and thus, the luminosity more difficult to determine. We also exclude sources with a Sersic profile effective radius $r_{e} < 0.05\arcsec$ to remove sources that might be stars or affected by the presence of a nearby star.

\begin{figure}[!ht]
\centering
\includegraphics[angle=90,width=3.4in]{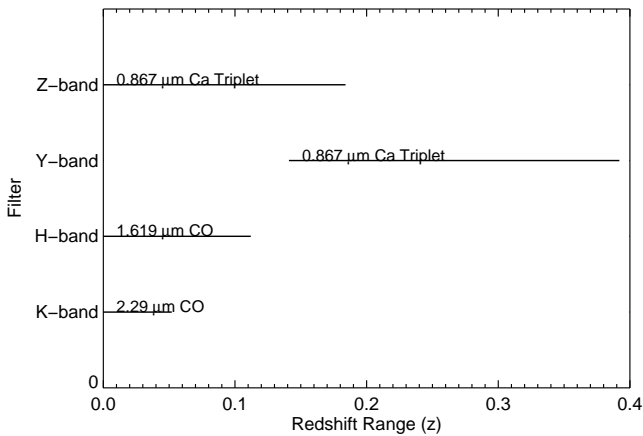}
\caption{The range of redshifts where different stellar spectral features will be observable with IRIS in different filters.}
\label{fig:redshift_compare}
\end{figure}

While the CO band-heads in \kband are ideal for measuring stellar dynamics, the strongest features are near the end of \kband at 2.2935 $\micron$ and 2.3227 $\micron$, which are shifted out of \kband at $z > 0.052$ and $z > 0.0384$, respectively. Beyond these distances, it is necessary to turn to other wavelengths for stellar features. In $H$ band, there are additional CO band-heads at 1.598, 1.619, and 1.640 $\micron$, which can also be used to measure line-of-sight velocity moments. These lines have smaller equivalent widths compared to the \kband CO band-heads, which means they require higher SNR to achieve the same precision for measurements of the kinematic moments compared to that of \kband. There are also more atmospheric OH emission lines in $H$ band that makes estimating the intrinsic line profiles more difficult \citep[see][]{2012ApJ...756..179M}. With $H$ band observations, the CO band-heads can be observed out to $z = 0.11$ before the strong 1.619 $\micron$ lines are redshifted out of the filter. Beyond this redshift, the $Z$ and $Y$ bands can be used to observe the Ca II triplet lines at 0.8498, 0.8542, and 0.8662 $\micron$. The 0.8662 $\micron$ line is redshifted out of \zband at $z = 0.184$. The Ca II triplet lines are redshifted into $Y$ band at $z$ = 0.141 (for the 0.8662 $\micron$ line) and out of $Y$ band at $z = 0.392$. For more details about using the Ca II triplet to measure the LOSVD, see Appendix \ref{append:zbb}. With the spectral-lines available in different filters, IRIS will be able to cover the redshifts of the majority of SDSS galaxies (see Figure \ref{fig:redshift_compare}). For the purposes of this study, we will limit ourselves to simulations at \kband and \zband, which constrains the redshift distribution to $0.005 < z < 0.184$ (luminosity distance of 21 Mpc to 746 Mpc).

\begin{figure}[th]
\centering
\includegraphics[width=3.4in]{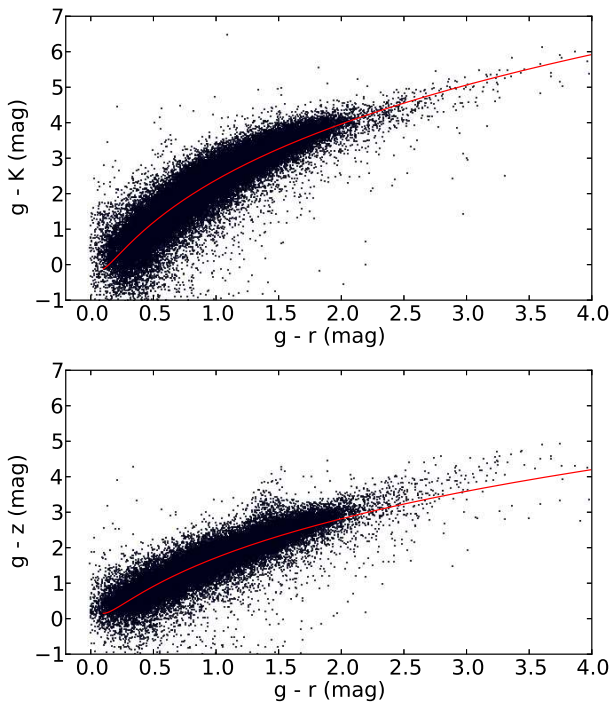}
\caption{The relationship between $g-r$ and $g - K$ colors (top), and $g-r$ and $g - z$ colors (bottom) for the GAMA galaxy sample \citep{2011MNRAS.413..971D}. We use these relationships to convert the surface brightness of SDSS galaxies observed at $r$ and $g$ to \kband and \zband. The red line is the best fit logarithm to the relationship between the two colors ($g-K = 2.0 \ln (g - r) + 0.396 (\ln (g-r))^2 + 2.387$, $g-K = 1.37 \ln (g - r) + 0.30 (\ln (g-r))^2 + 1.719)$.}
\label{fig:g_r_k}
\end{figure}

To convert between the $r$ and $g$ band surface brightness measurements from SDSS to \kband and \zband, we use the data from DR1 of the Galaxy and Mass Assembly survey \citep[GAMA,][]{2011MNRAS.413..971D}, which provides multi-wavelength observations from $u$-band to \kbandadj photometry of $\sim$120,000 SDSS selected galaxies. We fit a logarithmic function to the $g-r$ to $g-K$, and $g-z$ colors of the galaxies in order to find the conversion between $g-r$ to \kband and \zband, respectively (Figure \ref{fig:g_r_k}). Using the $g-r$ colors of the galaxy bulge measurements from \citet{2011ApJS..196...11S} and the relationship derived above, we scale the Sersic surface brightness profile fits and extrapolate these fits to obtain the $K$ and \zbandadj surface brightnesses at the centers of the galaxies for the simulations.

To estimate the predicted black hole mass in these galaxies, we use the $L-M_V$ relationship by converting the absolute M$_g$ and M$_r$ of the galaxy bulges to M$_V$ using the relationship $M_V = M_g - 0.533 (M_g - M_r) -0.00264$ \citep{1996AJ....111.1748F}. We then use the $M_V - \mbh$ relationship from \citet{2009ApJ...698..198G}, $\log{(M/M_\odot)} = 8.95 + 1.11\log{(L_V/10^{11} L_{\odot,V})}$, with an intrinsic root-mean squared scatter in $\log(\mbh/\msun)$ of $\epsilon_0 = 0.38$. As in Section \ref{sec:nearby_galaxies}, we use the inferred black hole mass to calculate the radius of influence of the black hole in the plane of the sky. This allows us to remove sources from our sample with a radius of influence less than 8 mas (18 mas), the spatial resolution of TMT at \zband (\kband). For all sources that pass these selection criteria, we compute the SNR of the observations at $R = 4000$ in the 4 mas (\zband) and 9 mas (\kband) plate scales with 2 hours of total integration time ($4 \times 900$ s). We define sources with peak SNR $> 40$ as those with black hole masses that will be measurable with IRIS on TMT, based on the typical SNR that are required today (see Section \ref{sec:current}). 

\begin{figure}[th]
\centering
\includegraphics[width=3.4in]{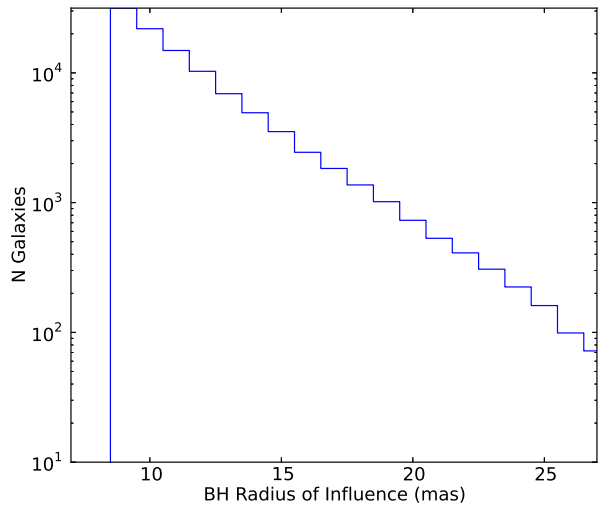}
\caption{The distribution of the gravitational radius of influence of black holes in SDSS galaxies that will be observable by IRIS using the \zband filter. The criteria for observability are detailed in Section \ref{sec:sdss}. The power-law nature of this distribution shows the tremendous impact of having access to high angular resolution observations at near optical wavelengths. The cut off at small radii is the 8 mas angular resolution limit of \zband with the 4 mas plate scale. About $10^5$ galaxies are observable at \zband; for comparison, about 4000 galaxies are observable at \kband (18 mas resolution).}
\label{fig:sdss_rinfluence}
\end{figure}

\begin{figure*}[th]
\centering
\includegraphics[width=6.5in]{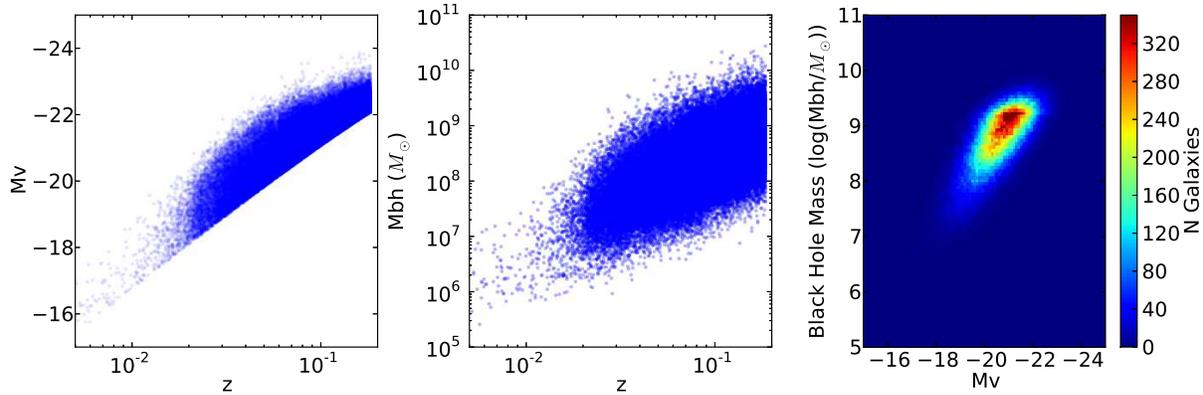}
\caption{The range of galactic nuclei available for dynamical mass estimates within the SDSS DR7 sample of galaxies. About $10^5$ galaxies from this sample are likely to be observable using the \zband, given their predicted black hole masses from the $M_V-L$ relationship and Sersic fits to their surface density profiles (see Section \ref{sec:sdss}). \textbf{Left:} the range of galaxy luminosities as a function of redshift. \textbf{Center:} The range in predicted black hole masses as a function of redshift. Scatter has been introduced into the masses according to the intrinsic scatter measured for the $M_V-L$ relationship measured by \citet{2009ApJ...698..198G}. \textbf{Right}: simulated $M_V-\mbh$ plot based on the sample of observable galaxies using the same scatter in BH masses. The density of galaxies are indicated by the colorbar. Black hole masses between $10^6-10^{10}$ \msun will be observable between $0.005 < z < 0.14$.}
\label{fig:sdss}
\end{figure*}

With these requirements, we find that IRIS on TMT will be able to measure black hole masses with stellar dynamics for about 4000 galaxies in the \kband and $10^5$ galaxies in the \zband from the SDSS DR7 sample. The difference in the number of accessible galaxies between \zband and \kband shows the tremendous power of increased angular resolution provided by access to the shorter wavelengths. Figure \ref{fig:sdss_rinfluence} show the distribution of the radius of influence of predicted black hole masses that will be accessible at \zband. These galaxies span a predicted mass range of $10^6-10^{10}$ \msun and a luminosity range of $-24 < M_V < -19$.  For comparison, there is currently a sample of about 72 galaxies with dynamical black hole mass measurements \citep[see recent compilation in][]{2013ApJ...764..184M}. The large sample of galaxies accessible with IRIS will enable an examination of the black hole scaling relation with different demographic samples of galaxies (Figure \ref{fig:sdss}). For example, recent analysis have suggested that the black hole relationships do not appear to apply to galaxies with either pseudobulges or bulge-less galaxies \citep{2011Natur.469..374K}; this may imply that the growth of black holes in these galaxies proceed differently than in galaxies with bulges, or that feedback mechanisms operate differently in these galaxies. Using the bulge to disk fraction, $B/T$ from \citet{2011ApJS..196...11S}, we find that the SDSS sample with accessible dynamical mass measurements span a wide range in $B/T$ from those that are disk-like with $B/T < 0.3$ to those that are completely bulge-dominated ($B/T > 0.75$); this sample should allow for much more detailed examinations of the origin of this discrepancy. Because of the intrinsic scatter in the black hole masses, at larger distances, a greater number of black holes will fall out of the observable sample (Figure \ref{fig:sdss}). It will be important for future surveys to account for this bias when selecting the range of black hole masses to study. For example, for $\mbh > 10^7 \msun$, this bias will set in at $z > 0.1$ (Figure \ref{fig:sdss}). 

\section{Comparison to current observations}
\label{sec:current}
In Table \ref{tab:prev_mbh}, we list some recent IFS measurements with their SNR, uncertainties on the Gauss-Hermite moments, Strehl ratios, and error in black hole mass. These observations are typically made using a plate scale of 50 mas, and the the spectra from multiple spaxels are generally binned in order to increase the SNR to about $\sim40$, with the highest SNR of about $\sim100$. The uncertainties in velocity dispersion are typically about 10 km s$^{-1}$ and in $h_3$ and $h_4$ about 0.02. From the simulations in Section \ref{sec:gauss_hermite}, we find that IRIS will be able to achieve comparable precision in the velocity moments at the same SNR. The uncertainty in black hole mass in Table \ref{tab:prev_mbh} varies over a large range, from 6\% to 50\%. The uncertainties in these measurements can be dominated by systematics relating to the dynamical models used to determine the black hole mass; in order to obtain an unbiased fit to the black hole mass from line-of-sight stellar kinematics, the dynamical models must simultaneously fit the black hole potential as well as the potential from the extended mass, such as the stellar distribution and dark matter halo. Each of these properties can have multiple parameters and assumptions (e.g. whether the stellar distribution is tri-axial or axisymmetric). The dependence of the black hole mass measurement on the complexities of the dynamical models introduces difficulty in estimating the uncertainty in the mass measurements from IRIS from the predicted SNR of the spectra alone. Nevertheless, current observations can inform us of the quality of the data from future IFSs that are necessary to achieve a precise black hole mass measurement. We expect that observations with IRIS will be able to achieve similar uncertainties in black hole masses as today, given the same analysis tools (but for a much wider range of measurements). Beyond just matching the SNR of current measurements, the much higher angular resolution measurements with TMT will be crucial to reduce the systematic effects of model assumptions. For example, in the case of M87, \citet{2009ApJ...700.1690G} showed that the mass of its central black hole increased by more than a factor of 2 when a dark matter halo is included in the models for data that poorly resolved the sphere of influence of the black hole \citep[see also,][]{2011ApJ...729...21S}. With higher angular resolution observations from the NIFS spectrograph on Gemini North, \citet{2011ApJ...729..119G} found that the inclusion of a dark matter halo into the dynamical model became less important and their final black hole mass measurement became much less sensitive the choice of model parameters. In the region where the kinematics are dominated by the potential of the black hole, the effect of the extended mass such as the star cluster and the bulge also become less important to the fit. 

The dynamical \mbh measurements from IRIS also have the potential to calibrate other methods of estimating black hole masses, such as reverberation mapping or single-epoch quasar spectra. These methods measure the virial masses by measuring the kinematics of the broad-line region (BRL) clouds around AGNs. However, the results depend strongly on assumptions about the geometry and kinematics of BRLs, as the measurement of the kinematics are not spatially resolved as in stellar dynamical measurements. Mass estimates using these methods have uncertainties of up to a factor of 3 \citep{2004ApJ...615..645O,2004ApJ...613..682P}. Stellar dynamical measurements of black hole masses can serve to better calibrate these methods, but only a small number of systems are currently suitable \citep{2006ApJ...646..754D,2007ApJ...670..105O}. IRIS should significantly increase the number of accessible sources. 

\subsection{Binning spatial locations}
A very common method employed in almost all current IFS dynamically measurements of black hole masses is the binning of several spatial locations in order to obtain a higher SNR measurement for the dynamical models. This binning is at the expense of spatial resolution, but without binning there is insufficient SNR to obtain robust dynamical constraints. Typically, the spaxels are binned less at the centers of the nuclei and more at the edges in order to obtain comparable SNR per bin. Most of the simulations in this paper of the capabilities of IRIS are made assuming either no binning or binning up to a resolution element. For example, in Table \ref{tab:nearby_early_type}, we tabulate both the peak SNR at a single spaxel as well as the integrated SNR binned up to the diffraction limit of 18 mas, which typically includes about 4 spaxels in the 9 mas plate scale and about 16 spaxels in the 4 mas plate scale. Further binning will be possible, as many sources will have many spaxels covering the angular size of the radius of influence in the sky. As today, the amount of binning will need to be optimized on an object by object basis in order to maximize both SNR and angular resolution. 

\begin{deluxetable*}{lcccccccccc}
\tablecolumns{12}
\tablecaption{Examples of previously reported BH masses from IFS observations}
\tablewidth{0pc}
\tabletypesize{\scriptsize}  
\tablehead{\colhead{Galaxy} & \colhead{Reference} & \colhead{Wavelength} & \colhead{Strehl Ratio} & \colhead{Binned SNR} & \colhead{BH Mass} & \colhead{$\Delta V$} & \colhead{$\Delta \sigma$} &  \colhead{$\Delta h_3$} & \colhead{$\Delta h_4$} & \colhead{Int. Time} \\
\colhead{} & \colhead{} &  \colhead{($\micron$)} & \colhead{} & \colhead{} & \colhead{(\msun)} & \colhead{(km s$^{-1}$)} & \colhead{(km s$^{-1}$)} & \colhead{} & \colhead{} & \colhead{(hr)} }
\startdata
M87   &  \citet{2011ApJ...729..119G} & \kband & 0.40 & 3-100 & $6.6\pm0.4 \times 10^9 $ &  7-18 &  9-20 & 0.02-0.04 &  0.01 - 0.03 &  3.8 \\
NGC 1332 & \citet{2011MNRAS.410.1223R} & \kband & 0.36 & 80-90 & $1.45\pm 0.20\times 10^9 $ &  8 & 9 &  0.02 &  0.02 &  0.7, 1.3 \\
NGC 6086 & \citet{2011ApJ...728..100M} & $H$ band & 0.25 & 20-40 & $3.6^{+1.7}_{-1.1}\times 10^9 $ &  40 &  26 &  0.042 &  0.024 &  2.25 \\
NGC 524 & \citet{2009MNRAS.399.1839K} & $H+K$ band & 0.31 &  40-60 & $8.3^{+2.7}_{-1.3}\times 10^8 $ &  10 &  13&  0.03&   0.04 &  3.5 \\
NGC 2549 & \citet{2009MNRAS.399.1839K}& $H+K$ band & 0.31 &  40-60 & $5.8^{+0.2}_{-1.3}\times10^7 $ &  5&  7 &  0.03&  0.03 &  2.5 \\
\enddata
\label{tab:prev_mbh}
\end{deluxetable*}

\section{Complementary Observations}
IRIS will achieve high SNR observations of the kinematics of nuclear regions, but other telescopes and instruments will be required to observe galaxies at larger scales to fully construct a dynamical model for black hole mass measurements. Measurements outside the nucleus helps to reduce systematic errors due to the distribution of the stellar light. For example, the light distribution affects the mass measurement through assumption about the distribution of the kinematic tracers and the mass-to-light ratio \citep[e.g.][]{2011ApJ...729..119G}. Deep spectroscopy from seeing-limited instruments such as VIRUS-P \citep{2013AJ....145..138B} or the Keck Cosmic Web Imager \citep{2010SPIE.7735E..21M} will provide essential kinematic information on the outer regions of nearby galaxies. The surface brightness profiles of these galaxies will ideally be provided by either the IRIS imager or \textit{James Webb Space Telescope} (\textit{JWST}).

\section{Conclusions}
\label{sec:conclusion}
The primary advantages of IRIS will be its ability to resolve the sphere of influence of the black hole in many systems with high SNR spectra. We find through simulations with realistic assumptions about the telescope and instrument that in most cases, IRIS will not only be able to achieve better angular resolution than current IFSs, it will be able to do so at least at comparable, if not higher SNR at much higher spatial sampling. For example, IRIS will be able to achieve the same SNR at the 9 mas spaxel scale as OSIRIS (the current IFS at Keck) at the 50 mas plate scale. With the $R = 8000$ mode, IRIS will also be able to definitively detect IMBHs of $\sim10^4$ \msun in the local group. Using the SDSS DR7 sample of galaxies, we find that IRIS will be capable of measuring black hole  masses for over $10^5$ galaxies, enabling the study of the demography of massive black holes at a level of detail not possible today. The number of observable galaxies are so vast that samples will need to be chosen carefully to obtain accurate BH demographics. Future observations will be able to define samples by scientific goals, rather than by observability as is often the case today.

Our current understanding of the relationship between central massive black holes and their host galaxies have reached a point where we are limited by the range of systems that are available for observations with the telescopes available today. The next leap in our understanding of the formation of central black holes and galaxy evolution will only be possible with the next generation of very large telescopes such as TMT in combination with the integral-field spectroscopic capabilities of an instrument like IRIS.

We thank the anonymous referee for helpful comments. The authors gratefully acknowledge the support of the TMT partner institutions. They are the Association of Canadian Universities for Research in Astronomy (ACURA), the California Institute of Technology and the University of California. This work was supported as well by the Gordon and Betty Moore Foundation, the Canada Foundation for Innovation, the Ontario Ministry of Research and Innovation, the National Research Council of Canada, Natural Sciences and Engineering Research Council of Canada, the British Columbia Knowledge Development Fund, the Association of Universities for Research in Astronomy (AURA), the U.S. National Science Foundation, and National Astronomical Observatory of Japan (NAOJ).  Research by A.J.B. is supported in part by NSF grant AST-1108835.

\appendix

\section{\zband Observations}
\label{append:zbb}

\begin{figure}[!ht]
\centering
\includegraphics[angle=90,width=6in]{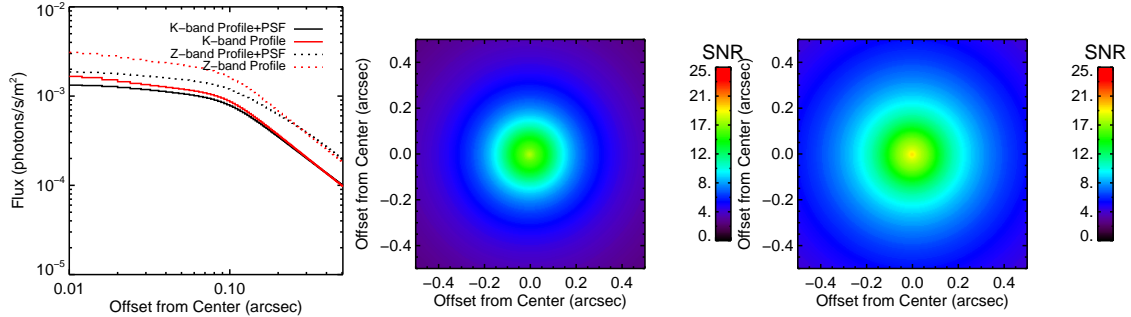}
\caption{Comparison of simulated observations in the \kband versus those in the \zband. The \zband will be able to provide higher spatial resolution, but at the expense of lower Strehls. \textbf{Left:} shows the best fit surface brightness profiles of NGC 4458 from \citet{2007ApJ...664..226L} for the \kband (solid) and the \zband (dotted) both before (red) and after (black) convolution with the PSF. The lower Strehl ratio in the \zband is apparent with the decrease in the central light concentration after accounting for the PSF. We use $V - K = 3.0$ and $V - Z = 1.42$ in order to convert the $V$-band colors in \citet{2007ApJ...664..226L} to the appropriate wavelengths. \textbf{Center:} the average SNR map for a single spectral channel in the \kband at the 4 mas plate scale, $R = 4000$, and an integration time of 900 s. \textbf{Right:} the average SNR in the \zband per spectral channel at the 4 mas plate scale, $R = 4000$, and an integration time of 900 s. The \zbandadj observations have higher SNR than at \kband, likely because of lower sky background.}
\label{fig:compare_kbb_zbb}
\end{figure}

One of the major aims of the AO system on TMT is to push AO corrections to shorter wavelengths. NIFRAOS will be able to deliver AO correction to \zband ($\lambda_\mathrm{central} = 0.928 \micron$), starting from 0.840 $\micron$ to 1.026 $\micron$. The on-axis Strehl ratios are predicted to be about 0.19 at zenith, which is comparable to the performance on today's telescopes at $J$ or $H$ bands. This will provide us with the potential to obtain up to $\sim8$ mas angular resolution. This wavelength also contains the Ca II triplets centered $\sim 0.85$ $\micron$, which is often used by STIS on \textit{HST} for observations to derive black hole masses. The \zband has also a further advantage of having much lower sky background compared to \kband with $Z_\mathrm{sky}$ = 18.9 mag compared to $K_\mathrm{sky}$ = 13.9 mag. In order to assess the potential advantage of \zband, we have simulated the same galaxy, NGC 4458 at these two wavebands from the \citet{2007ApJ...664..226L} sample. We simulate the \zbandadj observations using the 4 mas plate scale (the smallest plate scale, though undersampled) and the \kbandadj observations at the 4 mas plate scale (oversampled at \kband). Figure \ref{fig:compare_kbb_zbb} shows the SNR comparison between these two filters. The \zbandadj observations has slightly higher SNR compared to the \kbandadj observation due to lower sky background at \zband. Since the \kbandadj observations are oversampled, it would be possible to achieve higher SNR if binned to 9 mas to achieve Nyquist sampling at \kband. We also investigate the precision with which IRIS will be able to measure the moments of the LOSVD from the Ca II triplet of a 4500 K ($\sim$ K3III giant) template by using Monte Carlo simulations similar to our procedure with the CO lines in \kband. To simulate the effect of possible spectral template mismatch, we include stellar spectra with temperatures from 3000 to 4500 K in 100 K steps as possible templates for the routine to choose. We find that for a given SNR, the \zband has slightly higher precision in the velocity moments. 

One potential difficulty in using \zband for observations is that the lower Strehl may make the PSF more difficult to determine. An accurate PSF is crucial to be able to account for the effect of beam smearing on kinematic measurements \citep[e.g.][]{2009ApJ...699..421W}. It is unclear at this time how the uncertainty in the PSF will affect the conclusions in this work as we have not simulated PSF mismatch. While PSF knowledge is currently very limited in observations with the current generation of spectrographs, the GSMTs such as TMT will have the ability to reconstruct the instantaneous PSFs from AO telemetry information and atmospheric turbulence measurements \citep{2011aoel.confE..73G}. These techniques have the potential to mitigate many of the problems associated with PSF mismatch today.

\begin{figure}[!ht]
\centering
\includegraphics[width=4in]{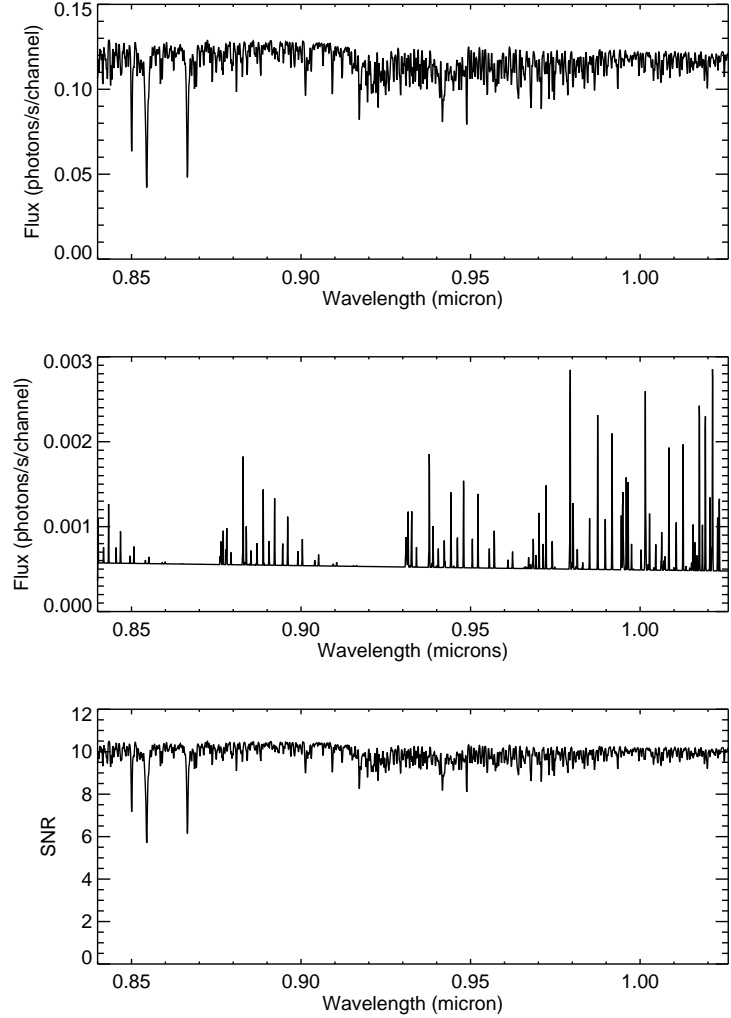}
\caption{Top: synthetic spectrum of a $\teff = 4500 K$ ($\sim$K3III) giant used for the simulations in a 4 mas spaxel in the \zband. The Ca II triplets are located at $0.8498$ $\micron$, 0.8542 $\micron$, and $0.8662$ \micron.The fluxes are given for observations of a location with surface brightness with $Z$ = 13 mag arcsec$^{-2}$,  at a spectral resolution $R = 8000$, and an integration time of 900 s. Middle: the background spectrum, which includes contributions from the sky continuum, telescope, OH lines, and zodiacal emission. Bottom: the resulting SNR at each spectral channel in a single spaxel.}
\label{fig:sample_snr_zbb}
\end{figure}

\begin{figure}[!ht]
\centering
\includegraphics[angle=90,width=6in]{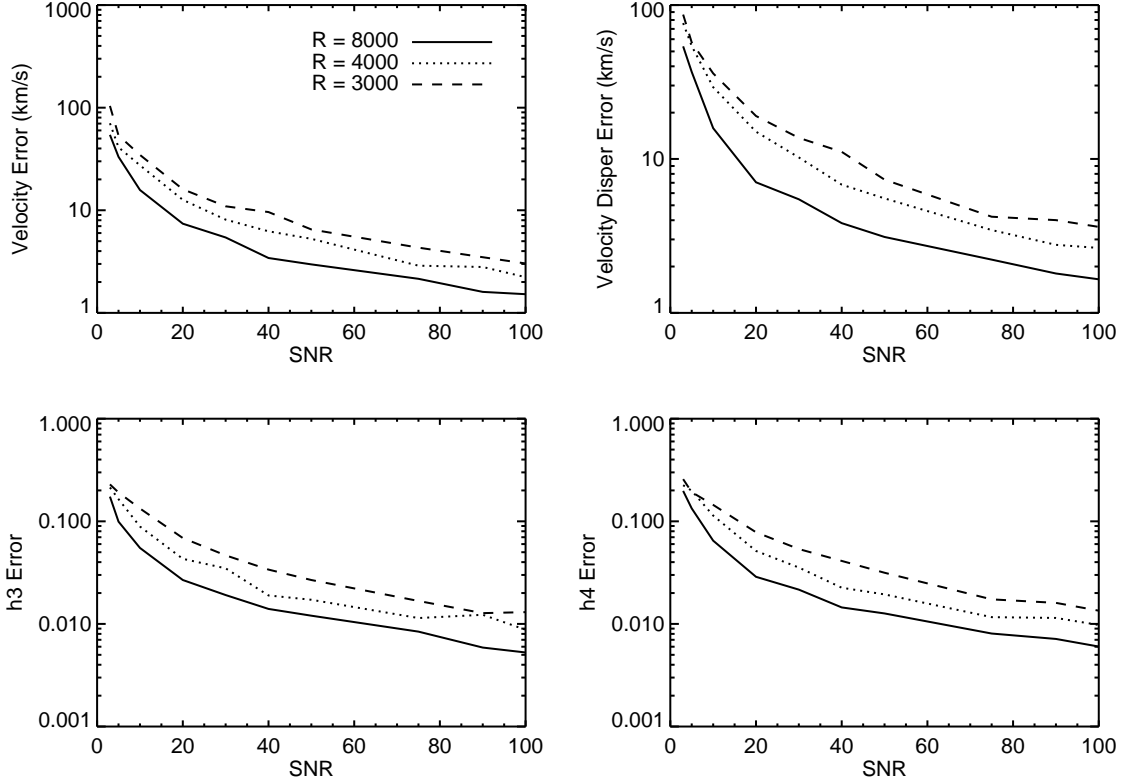}
\caption{Results of simulations of the dependence of the uncertainties in measured velocity moments on the SNR of the observed spectra at spectral resolutions of $R = 8000$ (solid) and $R = 4000$ (dotted) for observations in the \zband targeting the Ca II triplets. The simulations were run a synthetic spectrum of a K3III star, with Gauss-Hermite moments: $v$ = 0 km s$^{-1}$, $\sigma = 200$ km s$^{-1}$, $h_3 = -0.14$, and $h_4 = 0.03$.}
\label{fig:disp_err_sim_zbb}
\end{figure}

\bibliography{/Users/tdo/Documents/bibtex/prelim}


\end{document}

%% file: lauer_galaxy_sample_table_example.tex
NGC 3115&	 S0-  &	-21.1 &	11.5 &	10.2 &	0.89 &	0.52 &	89. &	394. &	196. &	392. &	1.9e+08 &	18.1 \\
NGC 3607&	 S0  &	-19.9 &	13.9 &	10.9 &	0.95 &	0.26 &	21. &	96. &	52. &	103. &	5.5e+07 &	9.0 \\
NGC 3384&	 S0-  &	-19.9 &	13.4 &	11.7 &	1.02 &	0.71 &	85. &	373. &	186. &	372. &	5.8e+07 &	8.6 \\
NGC 3377&	 E  &	-20.1 &	11.1 &	11.7 &	1.02 &	0.03 &	115. &	506. &	251. &	502. &	6.7e+07 &	9.3 \\
NGC 3379&	 E  &	-21.1 &	13.2 &	11.7 &	1.02 &	0.18 &	26. &	118. &	62. &	123. &	2.0e+08 &	16.0 \\
NGC 1023&	 S0-  &	-20.5 &	12.7 &	12.1 &	1.06 &	0.74 &	92. &	399. &	199. &	398. &	1.1e+08 &	11.3 \\
NGC 3056&	 S0+  &	-19.0 &	15.1 &	12.9 &	1.13 &	0.90 &	90. &	388. &	194. &	388. &	2.2e+07 &	4.8 \\